\documentclass[%
prd,nofootinbib,showpac,
superscriptaddress,
preprint
]{revtex4-1}
\usepackage{adjustbox}
\usepackage{graphicx}% Include figure files
\usepackage{dcolumn}% Align table columns on decimal point
\usepackage{bm}% bold math
\usepackage[utf8x]{inputenc} %for unicode character
\usepackage{amsmath,amssymb} %for equation
\usepackage{adjustbox}
\usepackage{color}

\usepackage[colorlinks,citecolor=blue]{hyperref}

\usepackage{float}
\usepackage{rotating}
\usepackage[font=small,labelfont=bf]{caption}
\usepackage{upgreek}%for a charactor in a equation
\usepackage{subcaption}

\newcommand{\dmatm}{$\Delta m^2_{31}$}
\newcommand{\absdmatm}{$|\Delta m^2_{31}|$}
\newcommand{\dmsol}{$\Delta m^2_{21}$}
\newcommand{\dcp}{$\delta_{\text{CP}}$}
\newcommand{\evmass}{$\times 10^{-3} \text{eV}^{2}/c^{4}$ }
\newcommand{\nova}{NO$\nu$A}
\newcommand{\nutwobb}{$\beta\beta_{2\nu}$}
\newcommand{\nulessbb}{$\beta\beta_{0\nu}$}

\begin{document}

\preprint{arXiv:xxxx.xxxx [hep-ph]}

\title{Neutrino mass spectrum: Present indication and future prospect}

%\thanks{Contact Author cvson@post.kek.jp}
	\author{S. Cao}
	\email{cvson@ifirse.icise.vn}
	\affiliation{\textit{Institute for Interdisciplinary Research in Science and Education,\\ \it{ICISE, Quy Nhon, Vietnam.}}}
	\affiliation{\textit{High Energy Accelerator Research Organization (KEK), Tsukuba, Ibaraki, Japan.}}
	
		\author{N. T. Hong Van}
%	\email{nhvan@iop.vast.ac.vn}
	\affiliation{\textit{Institute of Physics, Vietnam Academy of Science and Technology, Hanoi, Vietnam.}}
	\author{T. V. Ngoc}
    %\email{tranngocapc06@ifirse.icise.vn}
    \affiliation{\textit{Institute for Interdisciplinary Research in Science and Education,\\ \it{ICISE, Quy Nhon, Vietnam.}}}
	\affiliation{\textit{Graduate University of Science and Technology, Vietnam Academy of Science and Technology, Hanoi, Viet Nam.}}
	\author{P. T. Quyen}
    \affiliation{\textit{Institute for Interdisciplinary Research in Science and Education,\\ \it{ICISE, Quy Nhon, Vietnam.}}}
	\affiliation{\textit{Graduate University of Science and Technology, Vietnam Academy of Science and Technology, Hanoi, Viet Nam.}}

\date{\today}% It is always \today, today,

\begin{abstract}
The fact that neutrinos are massive has been the most crucial evidence of physics beyond the Standard Model of elementary particles. To date, we still do not know how neutrinos get mass and why their mass is much smaller than that of their charged fermion cousins. The precise determination of the neutrino mass spectrum has become one of the central tasks of neutrino physics, providing critical input for understanding the nature of neutrino mass and extending our model. The present landscape of the neutrino mass spectrum is reviewed and explored in this article using data from the neutrino oscillation, cosmology, and beta decay. In addition, we discuss the possibility of relevant programs elucidating the neutrino mass spectrum in the coming decades.
\end{abstract}

\maketitle

%\tableofcontents
\section{The Neutrino: An Elusive Misfit}
\begin{figure}[H]	
\centering
\includegraphics[width=0.7\textwidth]{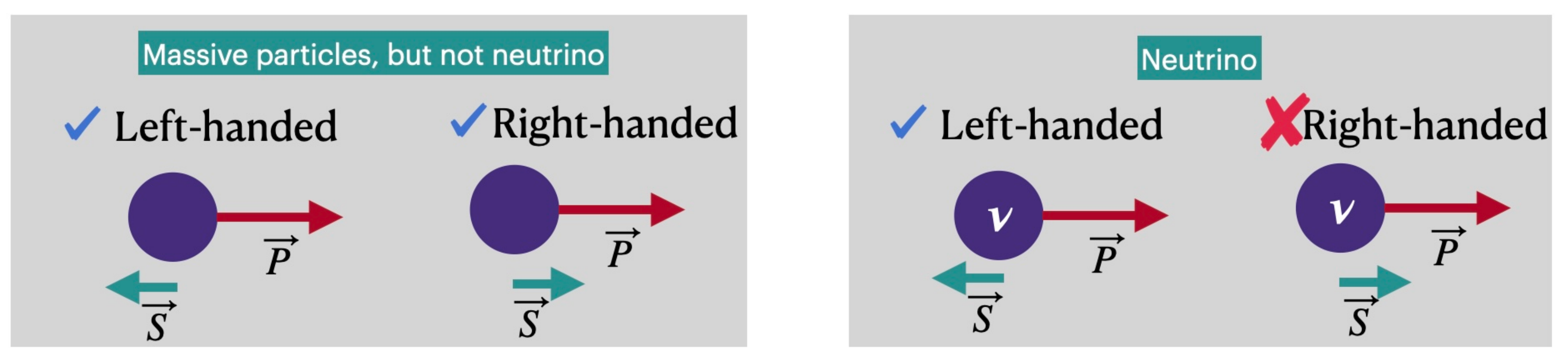}
\caption{The handiness of the neutrino compared to other massive particles A right-handed neutrino has not been found, and the neutrino is concluded to be massless in the Standard Model.\label{fig:nuparity}}
\end{figure}

We live in a matrix of neutrinos, which is the most abundant and possibly the most elusive of all the known massive particles. The neutrinos’ interactions dictate how the Sun shines, how the Sun evolves, and the dynamics of dying stars. In 1930, the existence of the neutrino was proposed by W. Pauli~\cite{Pauli:1930pc} as a desperate way out of interpreting the continuity of the beta particle's energy spectrum in the beta decay. However, it took 25 years for physicists to discover the neutrino fingerprint and obtain the neutrino's true birth certificate~\cite{Reines:1956rs}. 
The Standard Model (SM) of the elementary particle, our remarkably successful model for unifying three known interactions (electromagnetism, weak and strong interactions) predicts that neutrinos have a zero mass because all neutrinos are found to be left-handed~\cite{Goldhaber:1958nb}, as illustrated in Fig.~\ref{fig:nuparity}. However, the well-established neutrino oscillation~\cite{Zyla:2020zbs} indicates that neutrinos are massive and the leptons are mixed. As a result, neutrinos provide compelling evidence of physics beyond the SM. 

\noindent Although physicists place neutrinos in a special position on their study agenda, neutrinos remain among the most mysterious of all known particles to this day. In fact, we do not know yet how neutrinos gain mass or whether they are Dirac or Majorana particles. It is well-established that neutrino mass is extraordinarily small~\cite{Zyla:2020zbs} but what the absolute mass of neutrinos is and whether the neutrino mass spectrum underpins a specific hierarchy or fair-minded anarchy are unknown. We do not know for certain if the leptonic \emph{CP} violation exists in neutrino oscillations and how this violation, if discovered, relates to the baryon asymmetry in the Universe. Furthermore, physicists are ecstatic at the prospect of neutrinos breaking a fundamental law of nature, such as the Lorentz or \emph{CPT} symmetry, because neutrinos never cease to amaze us.

\noindent The article provides an observation on the neutrino mass spectrum. Section~\ref{sec:present} provides a summary of what we know about the neutrino mass spectrum from various sources, including neutrino oscillation, beta decay, undiscovered neutrino-less double beta decay, and cosmology. Section~\ref{sec:future} investigates the possibility of elucidating the neutrino mass spectrum in the coming decades. Before concluding, we discuss the hierarchical or anarchical textures underlying the neutrino mass matrix in Section~\ref{sec:discuss}. 

%%%%%%%%%%%%%%%%%%%%%%%%%%%%%%%%%%%%%%%%%%%%%%%%%%%
%%%%%%%%%%%%%%%%%%%%%%%%%%%%%%%%%%%%%%%%%%%%%%%%%%%
\section{Present Landscape of the Neutrino Mass Spectrum}\label{sec:present}
\begin{figure}[H]	
\centering
\includegraphics[width=0.7\textwidth]{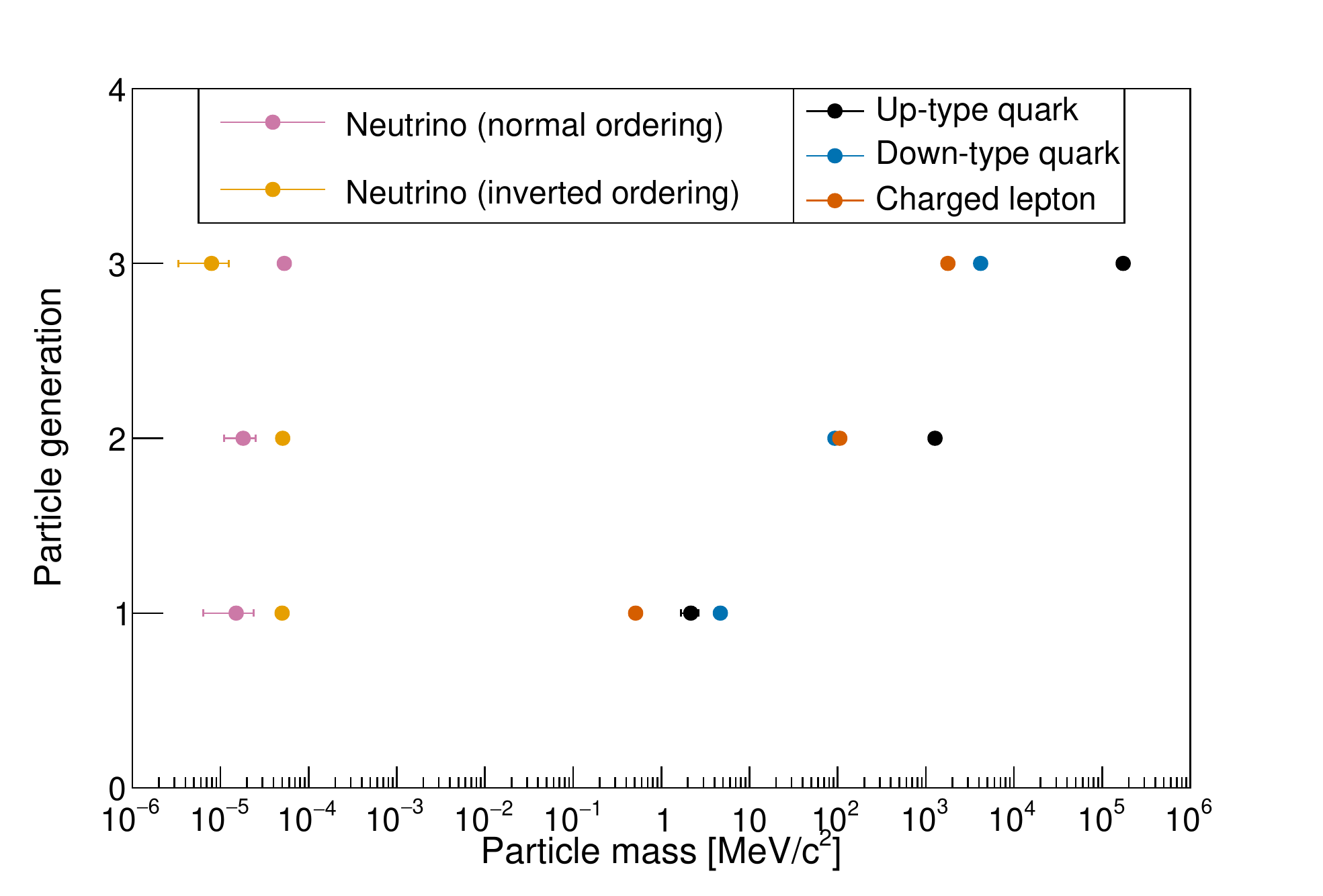}
\caption{Mass spectrum of all fermions. The quark and charged lepton mass are taken from PDG-2020~\cite{Zyla:2020zbs}. For the neutrino mass, constraints from the neutrino oscillation experiments, the cosmological constraint, and the effective electron neutrino mass from the beta decay are used. Neutrino mass is five order magnitude smaller than the electron mass.\label{fig:massspec}}
\end{figure}

Fig.~\ref{fig:massspec} depicts the mass spectrum of all fermions in the Standard Model. The mass spectrum of charged fermions (which include quarks and charged leptons) appears to be hierarchical.
\noindent One can read the mass texture via the dimension-less Yukawa coupling parameters
\begin{align*}
    \text{up-type quarks:} \ &y_{\text{u}} \sim 10^{-5},\ y_{\text{c}} \sim 10^{-2};\ y_{\text{t}} \sim 1; \\
    \text{down-type quarks:} \ 
    &y_{\text{d}} \sim 10^{-4}, \  y_{\text{s}} \sim 10^{-3}, \
    y_{\text{b}} \sim 10^{-2};\\
    \text{charged lepton:}\  &y_{e} \sim 10^{-6},\  y_{\mu} \sim 10^{-3};\
      y_{\tau} \sim 10^{-2}.\\ 
\end{align*}
The mass spectrum of charged fermions spans roughly six orders of magnitude. If neutrinos are included, the mass spectrum expands to more than eleven orders of magnitude. This extreme range of the fermion masses is one of the most perplexing aspects of the elementary particle world. 

The fermion mass spectrum clearly shows that neutrinos stand apart from others. The neutrino mass is roughly five orders of magnitude less than that of the smallest charged fermion particle, electron. Needless to say, weighing such an extraordinarily small mass as a neutrino is a difficult task. However, as the result of the tremendous efforts of numerous experiments over the last few decades, the neutrino mass spectrum is now more visible than ever. Our understanding of the neutrino mass spectrum is based on four major sources: (i) neutrino oscillation data, (ii) distortion of the beta-particle energy spectrum in the beta decay, (iii) cosmology data, and (iv) the search for the neutrino-less double beta decay. While the well-known neutrino oscillation phenomenon is direct evidence of neutrino massiveness, the data collected can only tell us about the absolute scale of the neutrino mass and is only sensitive to the mass square difference. To advance our knowledge of the neutrino mass, we must explore all the above-mentioned sources from a synergistic point of view.

%%%%%%%%%%%%%%%%%%%%%%%%%%%%%%%%%%%%%%%%%%%%%%%%%%%
%%%%%%%%%%%%%%%%%%%%%%%%%%%%%%%%%%%%%%%%%%%%%%%%%%%
\subsection{Neutrino mass from the neutrino oscillation}
\noindent Neutrino oscillation is a phenomenon where neutrino can change its flavor when given time to propagate. Neutrino oscillations necessitate the existence of neutrino mass spectrum, i.e mass eigenstates $\nu_i$ with definite mass $m_i$ (where $i$ is 1, 2, 3\footnote{Although the presently collected neutrino data samples support the three-flavor neutrino framework, it is still possible that there could be more than three mass eigenstates} at least). It requires flavor eigenstates with definite flavor, $\nu_{\alpha}$, where $\alpha$ are $e, \mu, \tau$, must be superposition of the mass eigenstates, \textit{a fundamental quantum mechanic phenomenon}. The relation between mass eigenstate $\nu_i$, flavor eigenstates $\nu_{\alpha}$ is expressed as
\begin{align}
    |\nu_{\alpha}\rangle = \sum_{i}U^*_{\text{PMNS},\alpha i}|\nu_{i}\rangle,
\end{align}
\noindent where $U_{\text{PMNS}}$ is a $3\times3$ unitary matrix. The matrix can be parameterized with three mixing angles $\theta_{12},\theta_{13}, \theta_{23}$ and one irreducible Dirac \emph{CP}-violation phase $\delta_{CP}$, formulated in a standard representation in Eq.~(\ref{eq:pmnsmatrix}). If the neutrino is a Majorana particle, there are two additional \emph{CP}-violation phases $\rho_1, \rho_2$, which play no role in neutrino oscillations.
%\textcolor{red}{Use ordering instead of hierarchy to avoid confusion of mass hierarchy in the Quark. }
\begin{align}\label{eq:pmnsmatrix}
      U_{\text{PMNS}}=
    \begin{pmatrix}
    c_{12}c_{13} & s_{12}c_{13} & s_{13}e^{-i\delta_{{CP}}}\\
    -s_{12}c_{23}-c_{12}s_{13}s_{23}e^{i\delta_{{CP}}} & c_{12}c_{23}-s_{12}s_{13}s_{23}e^{i\delta_{{CP}}} & c_{13}s_{23}\\
    s_{12}s_{23}-c_{12}s_{13}c_{23}e^{i\delta_{{CP}}} & -c_{12}s_{23}-s_{12}s_{13}c_{23}e^{i\delta_{{CP}}} & c_{13}c_{23}
    \end{pmatrix} \text{Diag}(e^{i\rho_{1}},e^{i\rho_{2}},0)
\end{align}
\noindent where $c_{ij} = \cos \theta_{ij}$ and $s_{ij} = \sin \theta_{ij}$. Neutrino oscillation is typically measured by comparing the flux of produced $\alpha$-flavor neutrinos with the flux of $\beta$-flavor neutrinos observed in a detector placed some distance from the production source. The probability for an $\alpha$-flavor to oscillate into $\beta$-flavor, $P_{({\nu_{\alpha}}\rightarrow\nu_{\beta})}$, depends on three mixing angles ($\theta_{12},\theta_{13},\theta_{23}$), \emph{CP}-violating phase \dcp, two mass-squared splittings ($\Delta m^{2}_{21}$, $\Delta m^{2}_{31}$) where $\Delta m^2_{ij}=m^2_{i}-m^2_{j}$, its energy $E_{\nu}$, propagation distance $L$, and the density of matter passed through by the neutrino $\rho$, given by
\begin{align}
P_{(\nu_{\alpha} \rightarrow \nu_{\beta})}=\delta_{\alpha \beta}&-4\sum_{i>j}Re(U^*_{\alpha i}U_{\beta i} U_{\alpha j} U^*_{\beta j})\sin^2 \left(\Delta m^2_{ij}\frac{L}{4E}\right) \nonumber\\
&\pm 2\sum_{i>j} Im(U^*_{\alpha i}U_{\beta i} U_{\alpha j} U^*_{\beta j})\sin \left(\Delta m^2_{ij}\frac{L}{2E}\right)
\end{align}
where $\Delta m^2_{ij} = m^2_i-m^2_j$. The $\pm$ stands for neutrino and anti-neutrino, respectively. By measuring the oscillation pattern/probability--typically as function of neutrino energy-- it enables us to unravel the neutrino mass-square differences $\Delta m^2_{21} = m^2_2-m^2_1$ and $\Delta m^2_{31} = m^2_3-m^2_1$, and determine the parameters of the leptonic mixing matrix. The global analysis of neutrino oscillation data is available e.g. in Ref.~\cite{esteban2019global,Esteban:2020cvm}, which is briefly summarized in Table~\ref{tab:nuoscpara}.
\begin{table}[h]
    \centering
    \begin{tabular}{|l|c|c|c|c|c|c|}
    \hline
    Parameter & $\sin^{2}\theta_{12}$ & $\sin^{2}\theta_{13}(\times 10^{-2})$ & $\sin^{2}\theta_{23}$ & $\delta_{\text{CP}}(^{\circ})$ &  $\Delta m^{2}_{21} (10^{-5}\text{eV}^{2})$ & $\Delta m^{2}_{31} (10^{-3}\text{eV}^{2})$\\
    \hline
    Best fit $\pm1\sigma$ & $0.310^{+0.013}_{-0.012}$ & $2.241^{+0.067}_{-0.066}$ & $0.558^{+0.020}_{-0.033}$ & $222^{+38}_{-28}$ & $7.39^{+0.21}_{-0.20}$ & $2.523^{+0.032}_{-0.030}$\\
   \hline
    \end{tabular}
    \caption{Global constraint of oscillation parameters with \emph{normal} neutrino mass ordering assumed, taken from  Ref.~\cite{esteban2019global,nufit5}. Here we assume $c=1$.}
    \label{tab:nuoscpara}
\end{table}
\noindent Regarding the order of the neutrino mass eigenvalues, we know that $m_2>m_1$ thank to the sizeable effect of the interaction between neutrino and the Sun's material, MSW effect~\cite{Wolfenstein:1977ue,Mikheyev:1985zog,Mikheev:1986wj}, on the survival of solar neutrinos~\cite{Borexino:2011ufb} observed in the neutrino experiments. With data from the solar neutrino experiments and KamLAND~\cite{esteban2019global,Esteban:2020cvm}, we now know that $\Delta m^2_{21}=2.74^{+0.21}_{-0.20}\times 10^{-5}$\evmass\ with a precision of 2.7$\%$. Evidently, $m_2\leq 0.005 \text{eV/}c^{2}$ but we do not know the absolute scale of $m_1$.

\begin{figure}[H]
\centering
\includegraphics[width=0.45\textwidth]{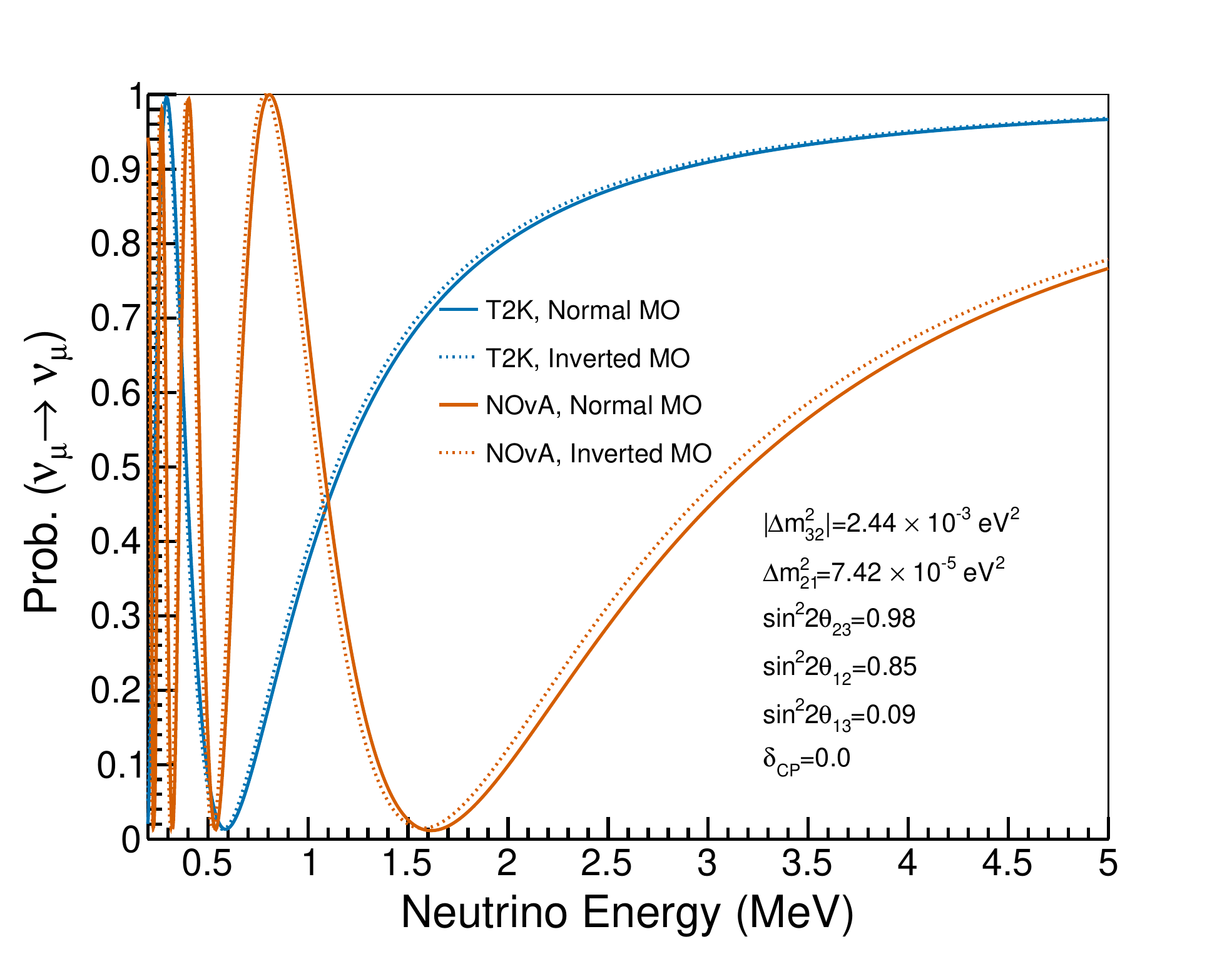}
\includegraphics[width=0.45\textwidth]{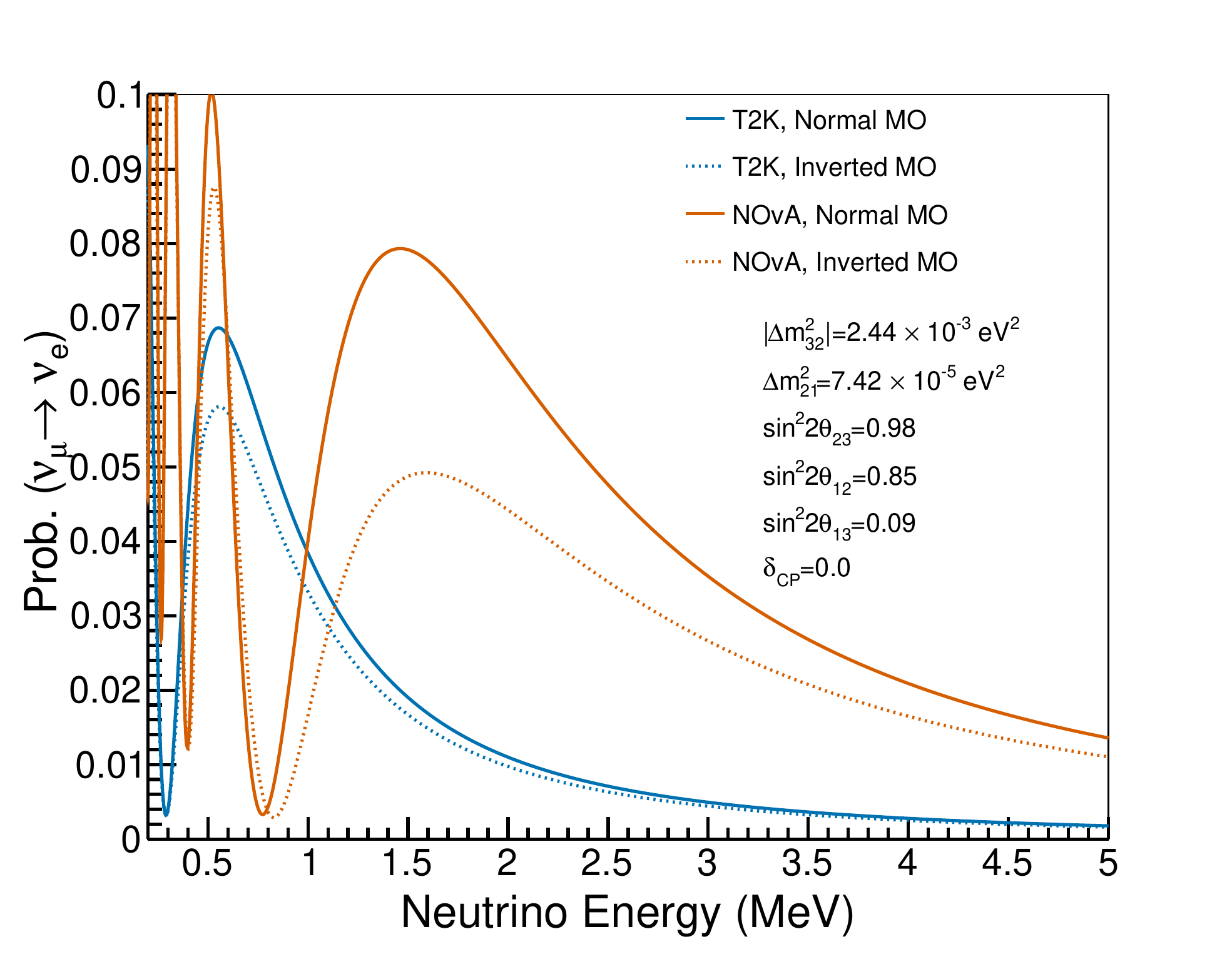}
\caption{Neutrino oscillation probabilities for $\nu_{\mu}\rightarrow\nu_{\mu}$ (or $\bar{\nu}_{\mu}\rightarrow \bar{\nu}_{\mu}$) disappearance (left) and $\nu_{\mu}\rightarrow\nu_{e}$ appearance (right) for the T2K and \nova\ experiments. For each experiment, two possible neutrino MO are considered.\label{fig:nuprob}}
\end{figure}

\noindent To get a full picture of the neutrino mass ordering, \dmatm\ needs to be measured precisely. In fact, thank to many experiments with the accelerator-based long-baseline neutrinos, atmospheric neutrinos, and reactor-based short-baseline neutrinos, we now know the amplitude of $|\Delta m^2_{31}|~\sim 2.5\times 10^{-3}$\evmass\ up to 1$\%$ precision. Since $\Delta m^2_{21}/|\Delta m^2_{31}|\approx 1\%$, practically we can assume $|\Delta m^2_{31}|\approx |\Delta m^2_{32}|\gg m^2_{21}$. However, the up-to-date neutrino data from these experiments can not tell conclusively whether $m_3>m_2>m_1$---known as the \emph{normal} mass ordering (MO)---or $m_2>m_1>m_3$---known as the \emph{inverted} MO.   Although T2K~\cite{patrick_dunne_2020_4154355}, \nova~\cite{alex_himmel_2020_4142045} and Super-Kamiokande~\cite{yasuhiro_nakajima_2020_4134680} experiments report individually that their data favor mildly the \emph{normal} neutrino MO, some studies, such as Ref.~\cite{Kelly:2020fkv}, show no favor in that indication with the combined data from the experiments. This ambiguity in determining the neutrino MO is worthy of further investigation. Measurement of \absdmatm\ is carried on in three main chain channels: (i) $\nu_{\mu}\rightarrow\nu_{\mu}$ (or $\bar{\nu}_{\mu}\rightarrow \bar{\nu}_{\mu}$ ) with accelerator-based long-baseline (A-LBL) neutrino experiments and atmospheric neutrino experiments; (ii) $\nu_{\mu}\rightarrow\nu_{e}$ (or $\bar{\nu}_{\mu}\rightarrow \bar{\nu}_{e}$ ) with A-LBL neutrino experiment; and (iii) $\bar{\nu}_{e}\rightarrow \bar{\nu}_{e}$ with the reactor-based short-baseline (R-SBL) neutrino experiments. The probabilities for two cases \emph{normal} and \emph{inverted} neutrino MO as function of the neutrino energies for the first two channels are shown in Fig.~\ref{fig:nuprob}.

\begin{figure}[H]
\centering
\includegraphics[width=0.45\textwidth]{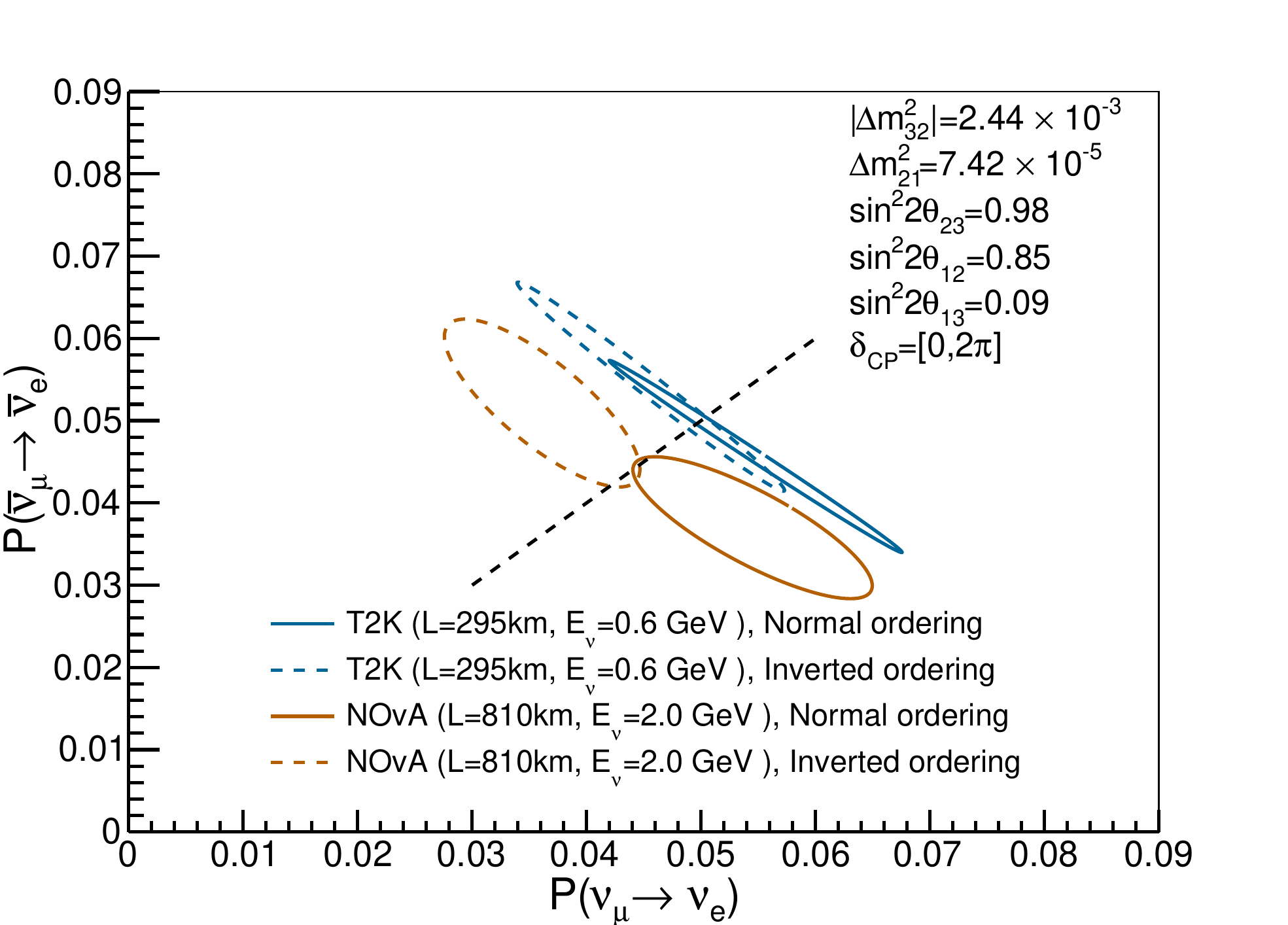}
\includegraphics[width=0.45\textwidth]{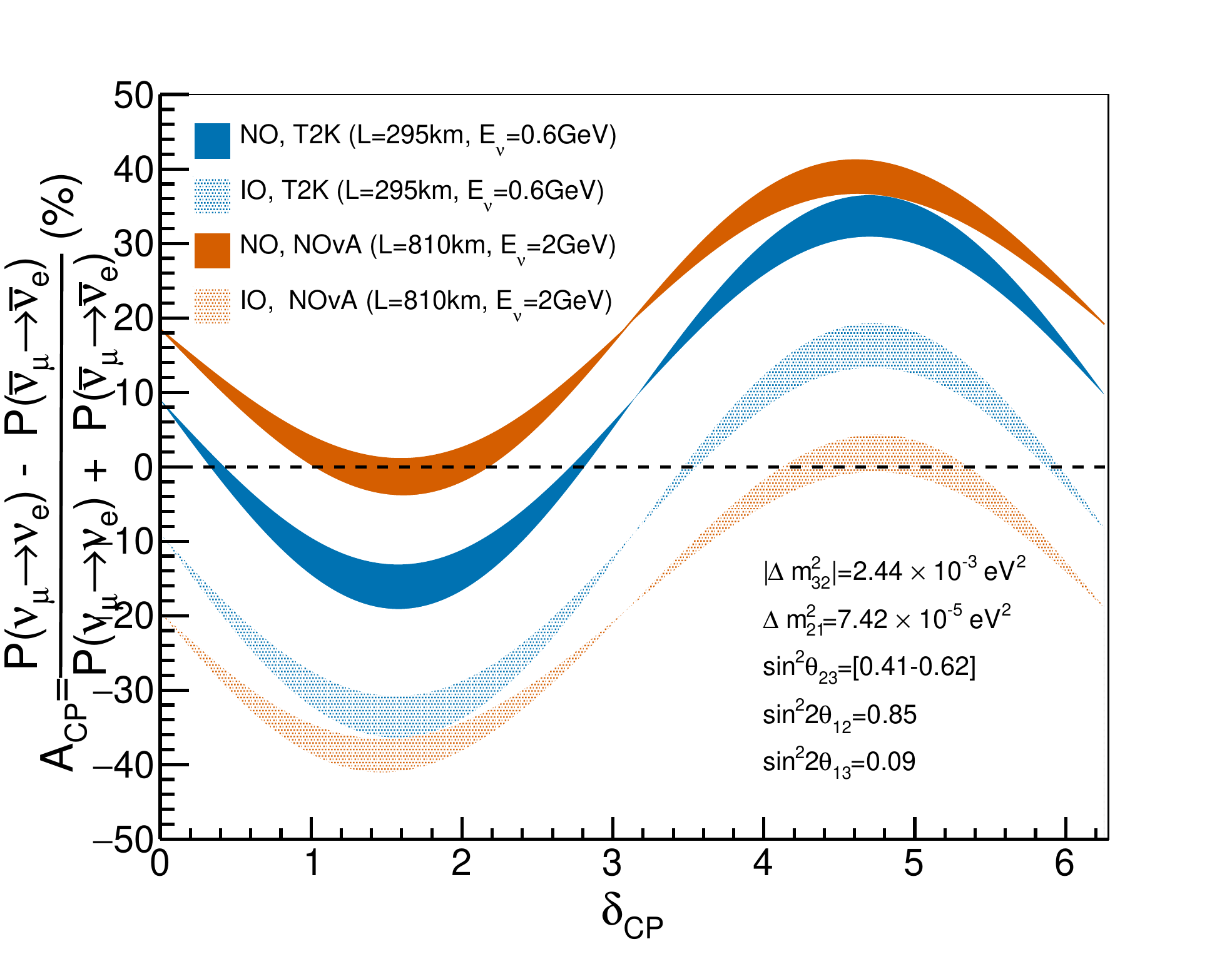}
\caption{For the T2K and \nova\ experiments, the bi-probability of $(\nu_{\mu}\rightarrow \nu_e)$ and $(\bar{\nu}_{\mu}\rightarrow \bar{\nu}_e)$ oscillation probabilities is shown on the left. The $A_{\text{CP}}$ quantity as function of $\delta_{CP}$ is shown on the right. The left graph is plotted at a single value of the mixing angle $\theta_{23}$, whereas the right takes the uncertainty of $\theta_{23}$ into account and is presented by the solid band in each graph.\label{fig:acpvsmassh}}
\end{figure}

\noindent It is pronounced that the sensitivity to the neutrino MO is marginal in the $\nu_{\mu}\rightarrow\nu_{\mu}$ (or $\bar{\nu}_{\mu}\rightarrow \bar{\nu}_{\mu}$ disappearance channels. The effect of the neutrino MO, on the other hand, is much stronger in the $\nu_{\mu}\rightarrow\nu_{e}$ (or $\bar{\nu}_{\mu}\rightarrow \bar{\nu}_{e}$ appearance channels. The relatively large modification of the oscillation probabilities in this channel is due to the coherent scattering of electron (anti-) neutrinos on the electrons present in the matter---the MSW effect. However, one must consider the fact that the $\nu_{\mu}\rightarrow\nu_{e}$ (or $\bar{\nu}_{\mu}\rightarrow \bar{\nu}_{e}$) appearance probability is just in few percentage,limiting the statistics of the collected data sample. Also, extracting the neutrino MO effect from the appearance probabilities is non-trivial since the sign of \dmatm\ is tangled severely with \dcp\ and the mixing angle $\theta_{23}$\footnote{It also depends on the mixing angle $\theta_{13}$, which we know with 3$\%$ precision}, which have been measured with relatively large uncertainty. Besides, it is important to note that the modification of the $\nu_{\mu}\rightarrow\nu_{e}$ and  $\bar{\nu}_{\mu}\rightarrow \bar{\nu}_{e}$ appearance probabilities due to matter effect is not the same. The left of Fig.~\ref{fig:acpvsmassh} shows the bi-probability of the $\nu_{\mu}\rightarrow\nu_{e}$ oscillation and $\bar{\nu}_{\mu}\rightarrow \bar{\nu}_{e}$ oscillation with T2K and \nova\ experiments. The plots are obtained at the best-fit value of the mixing angle $\theta_{23}$ and for two possible neutrino MO.  The bi-probability plot shows a better distinction between \emph{normal} and \emph{inverted} neutrino MO with \nova than with T2K. It is primarily due to the difference in the experimental baseline. The longer the experimental baseline is the larger the effect on the $\nu_{\mu}\rightarrow\nu_{e}$ and  $\bar{\nu}_{\mu}\rightarrow \bar{\nu}_{e}$ appearance probabilities. The sensitivity of the A-LBL experiments like T2K and \nova\ to \dcp\ and neutrino MO can be conceivable via the following expression of the so-called \emph{CP} asymmetry~\cite{Suekane:2015yta}, which presents a relative difference between $P_{(\nu_{\mu}\rightarrow \nu_e)}$ and $P_{(\bar{\nu}_{\mu}\rightarrow \bar{\nu}_e)}$ near the oscillation maximum, which corresponds to $\frac{|\Delta m^2_{31}|L}{4E_{\nu}}\approx \pi/2$. 
 \begin{equation} \label{eq:adcp} 
    A_{\text{CP}}\equiv \frac{P_{(\nu_{\mu}\rightarrow \nu_e)}- P_{(\bar{\nu}_{\mu}\rightarrow \bar{\nu}_e)}}{P_{(\nu_{\mu}\rightarrow \nu_e)} + P_{(\bar{\nu}_{\mu}\rightarrow \bar{\nu}_e)}}\Bigg|_{\frac{|\Delta m^2_{31}|L}{4E_{\nu}}\approx \pi/2}   \sim -\frac{\pi \sin 2\theta_{12}}{\tan\theta_{23}\sin 2\theta_{13}} \frac{\Delta m^2_{21}}{|\Delta m^2_{31}|} \sin \delta_{\text{CP}} \pm \frac{L}{2800km},
\end{equation}
 \noindent where $+(-)$ sign is taken for the \emph{normal} (\emph{inverted}) MO respectively. With values listed in Table~\ref{tab:nuoscpara}, $\frac{\pi \sin 2\theta_{12}}{\tan\theta_{23}\sin 2\theta_{13}} \frac{\Delta m^2_{21}}{|\Delta m^2_{31}|} \sim 0.256$, which means the \emph{CP} violation effect can be observed somewhat between $-25.6\%$ and $+25.6\%$. For a 295~km baseline of the T2K experiment, MO effect is subdominant with $\sim 10.5\%$.  This is illustrated in Fig.~\ref{fig:acpvsmassh}
\noindent Due to the mutual dependence of the considering parameters manifested in the neutrino oscillation probabilities, determining the neutrino MO will apparently enhance the sensitivity of the \emph{CP}-violation search and vice versa. As a result, the program to elucidate the neutrino MO and the search for \emph{CP} violation in the A-LBL neutrino experiments are inextricably linked.

\begin{figure}[H]
\centering
\includegraphics[width=0.45\textwidth]{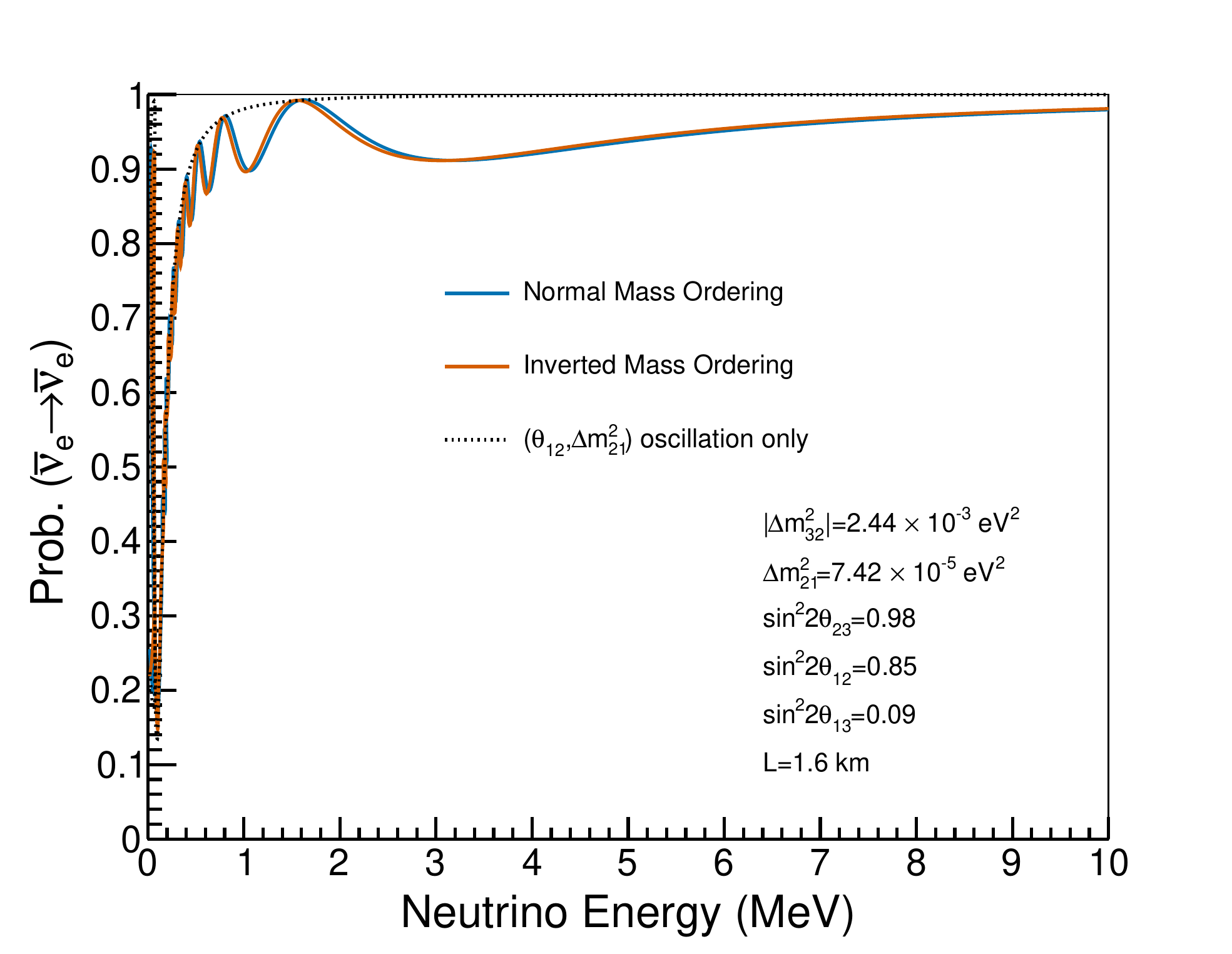}
\includegraphics[width=0.45\textwidth]{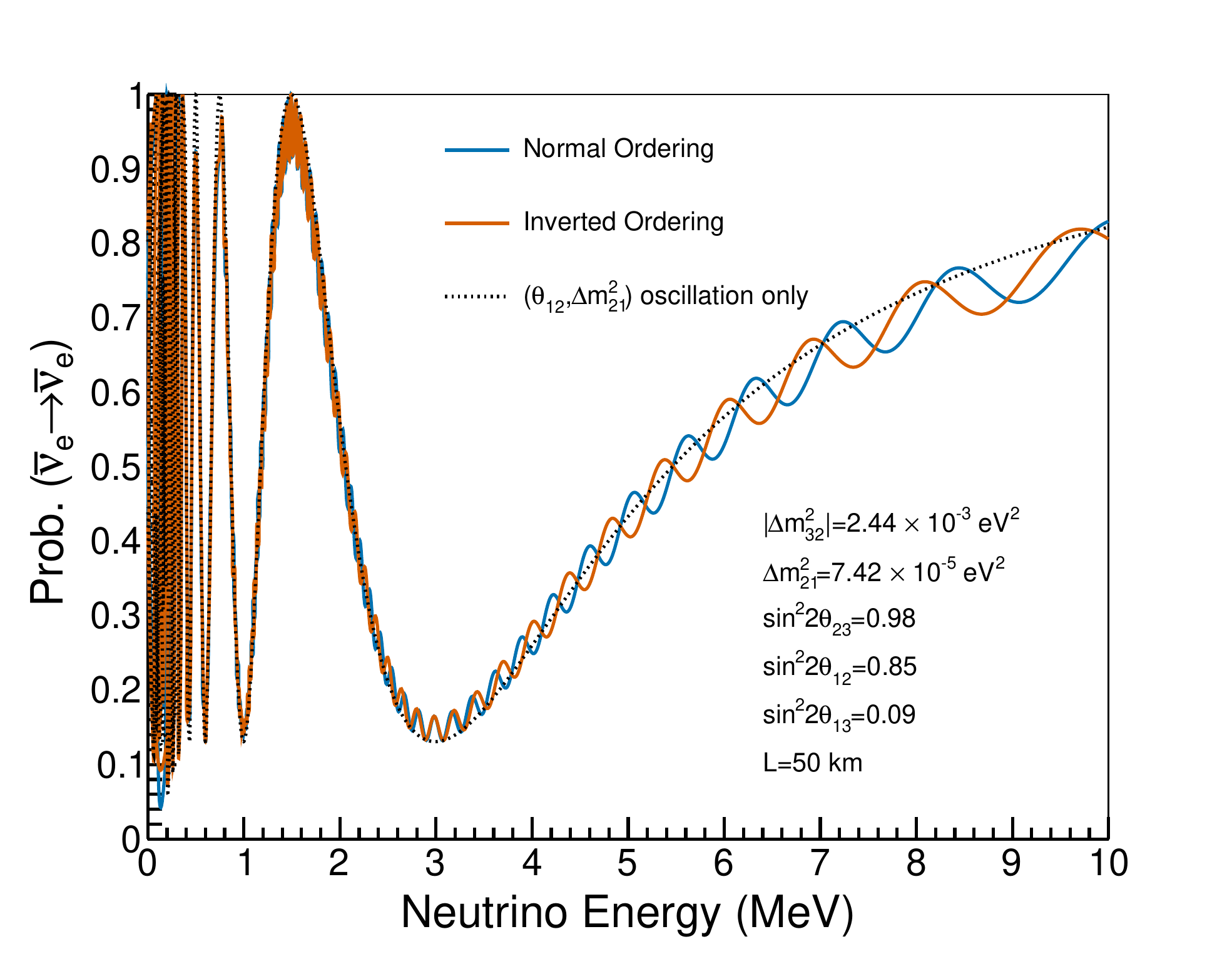}
\caption{The $\bar{\nu}_{e}\rightarrow \bar{\nu}_{e}$  oscillation probabilities in the reactor-based neutrino experiment with short (left) and medium (right) experimental baselines.\label{fig:nuprobreact}}
\end{figure}

So far, in the case of the reactor-based neutrino experiments, we have investigated with detectors placed relatively close to the reactor core, a few hundred meters to a few kilometers from the neutrino source. As shown on the left side of Fig.~\ref{fig:nuprobreact}, the sensitivity to the neutrino MO is marginal. However, a brand-new reactor-based experiment, JUNO~\cite{djurcic2015juno}, with a medium-baseline of 50~km, can improve the sensitivity to the neutrino MO thanks to the interference between two oscillation terms~\cite{petcov2002lma}, which are driven by \dmsol and \dmatm, respectively.
\noindent For JUNO and the like, the most challenging thing is to achieve an excellent resolution of the reconstructed neutrino energy for unravelling the neutrino MO effect from the detector response effect. The recent progress~\cite{Abusleme:2020lur} on the JUNO calibration demonstrates that this unprecedented achievement in energy resolution is viable. Besides, it is important to note that, unlike when measuring the neutrino MO with an A-LBL experiment, the sensitivity to the neutrino MO in the R-MBL neutrino experiments is independent of the value of \dcp. Thus, resolving the neutrino MO with A-LBL and R-MBL neutrino experiments is complementary~\cite{Cao:2020aig}.

%The limit of the experimental sensitivity to the neutrino mass ordering can be understandable from the $\nu_{\mu} (\bar{\nu_{\mu}})\rightarrow\nu_{\mu}(\bar{\nu_{\mu}})$ probability measured by accelerator-based and atmospheric neutrino experiments, and the $\bar{\nu}_e\rightarrow\bar{\nu}_e$ probability measured by the reactor-based experiment, formulated in Eq.~\ref{eq:nuebdis}.

%%%%%%%%%%%%%%%%%%%%%%%%%%%%%%%%%%%%%%%%%%%%%%%%%%%
%%%%%%%%%%%%%%%%%%%%%%%%%%%%%%%%%%%%%%%%%%%%%%%%%%%
\subsection{Neutrino mass constraint from beta decay and cosmology}
As pointed out in Ref.~\cite{curran1948beta}, the shape of the $\beta$-particle energy spectrum near its upper limit in the beta-decay process is extremely sensitive to the rest mass of the neutrinos. The decay,
\begin{align*}
    %^{3}H \rightarrow ^{3}He+e^{-}+\bar{\nu_i},\  \text{where}\ i=1,2, \text{or}\ 3.
    (A,Z) \rightarrow (A,Z+1)+e^{-}+\bar{\nu_i},\  \text{where}\ i=1,2, \text{or}\ 3.
\end{align*}
The three processes with three $\bar{\nu_i}$ respectively contribute incoherently to the total decay rate. As a consequence, the modification of the $\beta$-particle energy spectrum can be characterized by an effective neutrino mass given as
\begin{align*}
    \langle m_{\nu_{\beta}}\rangle = \sqrt{|U_{e1}|^2m_{\nu_{1}}^2 + |U_{e2}|^2m_{\nu_{2}}^2 + |U_{e3}|^2m_{\nu_{3}}^2}.
\end{align*}
Since the neutrino is extremely tiny, essential elements for this kind of measurement are a very high-resolution spectrometer---at the level of the eV to sub-eV resolution--- and a highly statistical data sample. Recently, the KATRIN experiment~\cite{Aker:2021gma} reported the measurement with sub-eV sensitivity and found that
\begin{align*}
    \langle m_{\nu_{\beta}}\rangle  <0.8 \text{ eV/c} \text{ at 90$\%$ C.L.}
\end{align*}
Another possibility to examine the absolute neutrino mass scale comes from the search for the so-called neutrino-less double beta (\nulessbb) decay. The process is similar to the double beta decay (\nutwobb) except that the former process has no neutrino produced in the final state,
\begin{align*}
    \text{Double beta decay $\beta\beta_{2\nu}$:}\ &(A,Z)\rightarrow (A,Z+2) +2e +2\bar{\nu_e},\\
    \text{Neutrinoless double beta decay $\beta\beta_{0\nu}$:}\ &(A,Z)\rightarrow (A,Z+2) +2e.
\end{align*}
If the neutrino is a Majorana particle, the eigenstates of the neutrino and anti-neutrino have no quantum number to distinguish them and can annihilate each other when they meet. This annihilation can be applicable to two neutrinos produced virtually in the double beta decay. A definite discovery of the \nulessbb\ will confirm the Majorana nature of the neutrino and reveal that the lepton number conservation is violated. The rate of this exotic \nulessbb\ decay is proportional to the so-called effective Majorana neutrino mass quantity, written as
\begin{align*}
\langle m_{\beta\beta}^{0\nu}\rangle = \sum_i U_{ei}^2 m_{\nu_i} = |m_{\nu_1}\cos^2\theta_{13} \cos^2\theta_{12} + m_{\nu_2}\cos^2\theta_{13} \sin^2\theta_{12}e^{i\rho_1}+m_{\nu_3}\sin^2\theta_{13}e^{i\rho_2}|,
\end{align*}
where $\rho_1,\rho_2$ are two Majorana phases, and it can take from 0 to $\pi$ as argued in Ref.~\cite{deGouvea:2008nm}. 

\noindent The recent constraint on the effective Majorana mass $\langle m_{\beta\beta}^{0\nu}\rangle$ comes from the Kamland-Zen experiment~\cite{KamLAND-Zen:2016pfg}. The data established an upper limit for the effective Majorana mass to be 
\begin{align*}
\langle m_{\beta\beta}^{0\nu}\rangle<61-165 \text{ meV/c} \text{ at 90$\%$ C.L.}
\end{align*}

Another probe for determining the absolute mass scale of neutrinos comes from cosmology. Since neutrinos are the second most abundant particles in the Universe, the aggregated neutrino mass leads to sizeable impacts on the cosmic microwave background and the large-scale structure~\cite{wong2011neutrino}. Recent results from Planck and Baryon Acoustics Oscillation (BAO) ~\cite{Aghanim:2018eyx} yield a striking constraint on the sum of the neutrino mass, 
\begin{align*}
\sum_{N_{\nu}} m_{\nu_i}=m_{\nu_{1}} + m_{\nu_{2}} + m_{\nu_{3}} <0.12 \text{ eV/c} \ \text{ at 90$\%$ C.L.}
\end{align*}
 \noindent Neutrino mass is one of the great examples of using cosmology to explore particle physics. 

\begin{figure}[H]	
\centering
\includegraphics[width=0.47\textwidth]{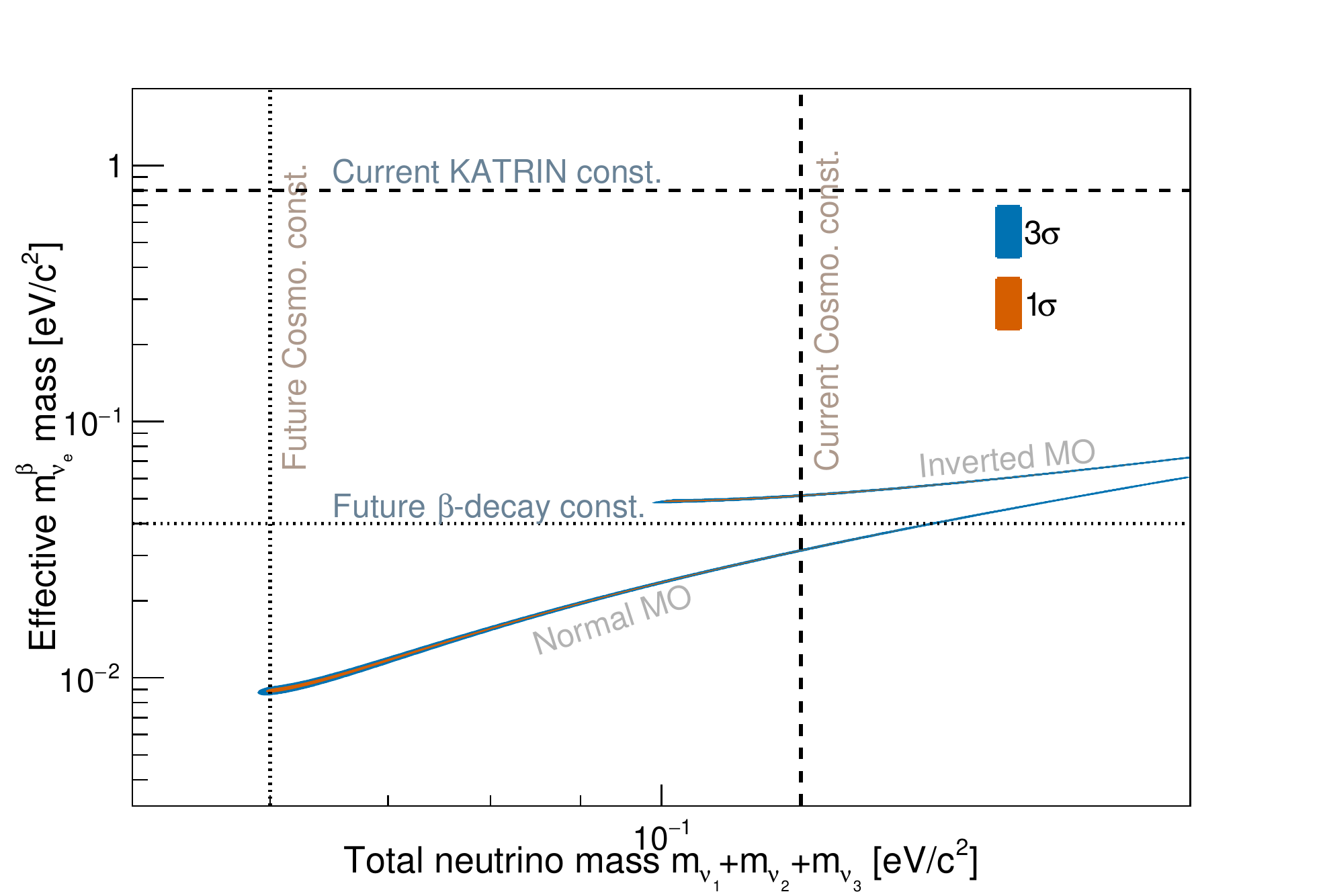}
\includegraphics[width=0.47\textwidth]{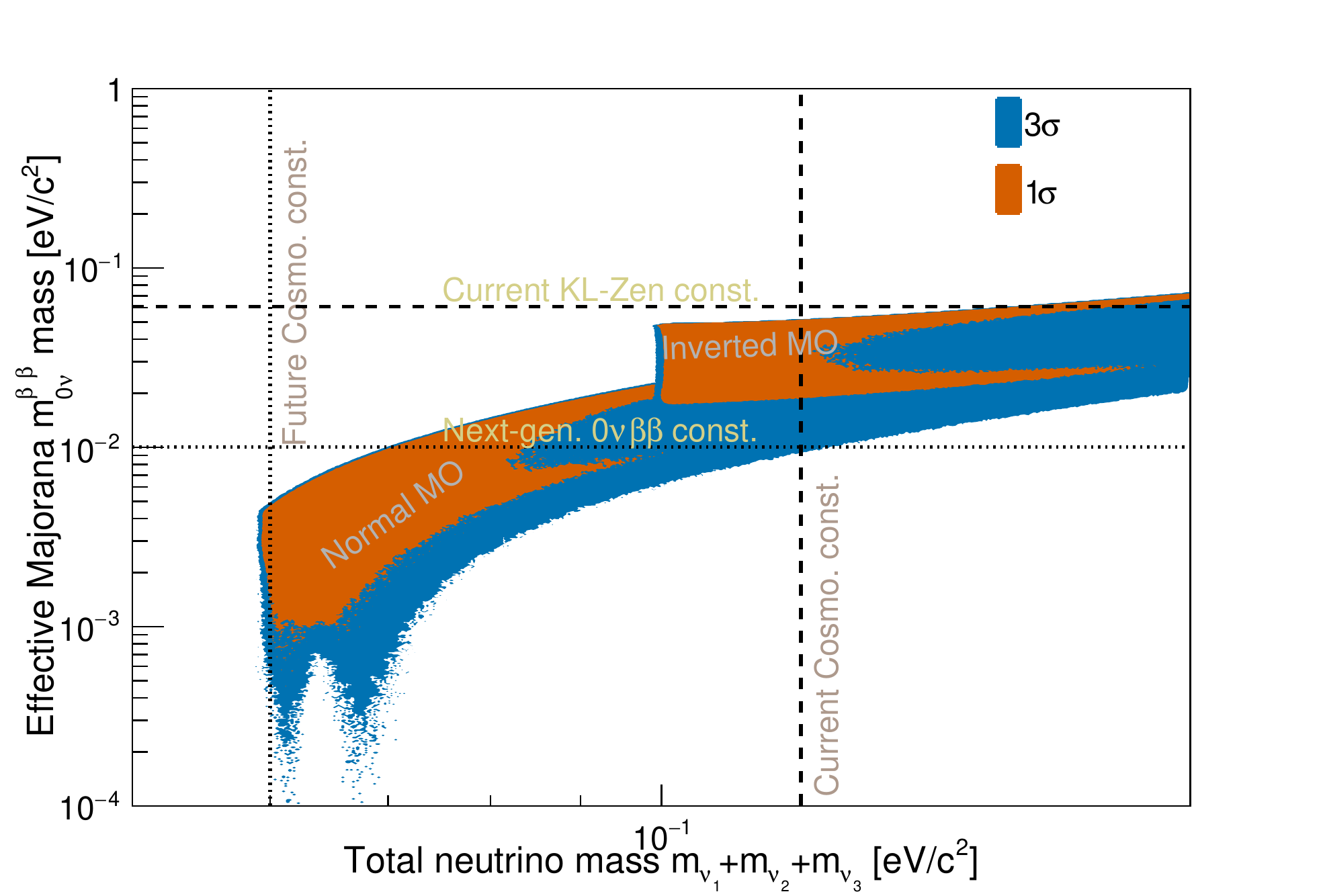}
\caption{Allow regions of $\sum_{N_{\nu}} m_{\nu_i}$, effective $\langle m_{\nu_{\beta}}\rangle$, and effective Majorana mass $\langle m_{\beta\beta}^{0\nu}\rangle$  computed from the latest global constraint~\cite{nufit5} of the neutrino mass-square difference obtained from the neutrino oscillation data. The present and future constraints from the beta-decay, the search for the neutrino-less double beta decay, and cosmology are included.\label{fig:massconstraint}}
\end{figure}

From the neutrino oscillation data, one can use constraints on the neutrino mass-square splittings \dmsol\, \dmatm\ and the leptonic mixing parameters to compute the distribution of the observable quantities of the absolute neutrino mass scales, including $\sum_{N_{\nu}} m_{\nu_i}$, effective $\langle m_{\nu_{\beta}}\rangle$, and effective Majorana mass $\langle m_{\beta\beta}^{0\nu}\rangle$. The results are shown in Fig.~\ref{fig:massconstraint}.
\noindent It is noteworthy that the present constraints on the absolute neutrino mass scale do not allow us to determine definitively the neutrino MO.

%%%%%%%%%%%%%%%%%%%%%%%%%%%%%%%%%%%%%%%%%%%%%%%%%%%
%%%%%%%%%%%%%%%%%%%%%%%%%%%%%%%%%%%%%%%%%%%%%%%%%%%
\subsection{A reconstructive neutrino mass matrix}
\begin{figure}[H]	
\includegraphics[width=0.47\textwidth]{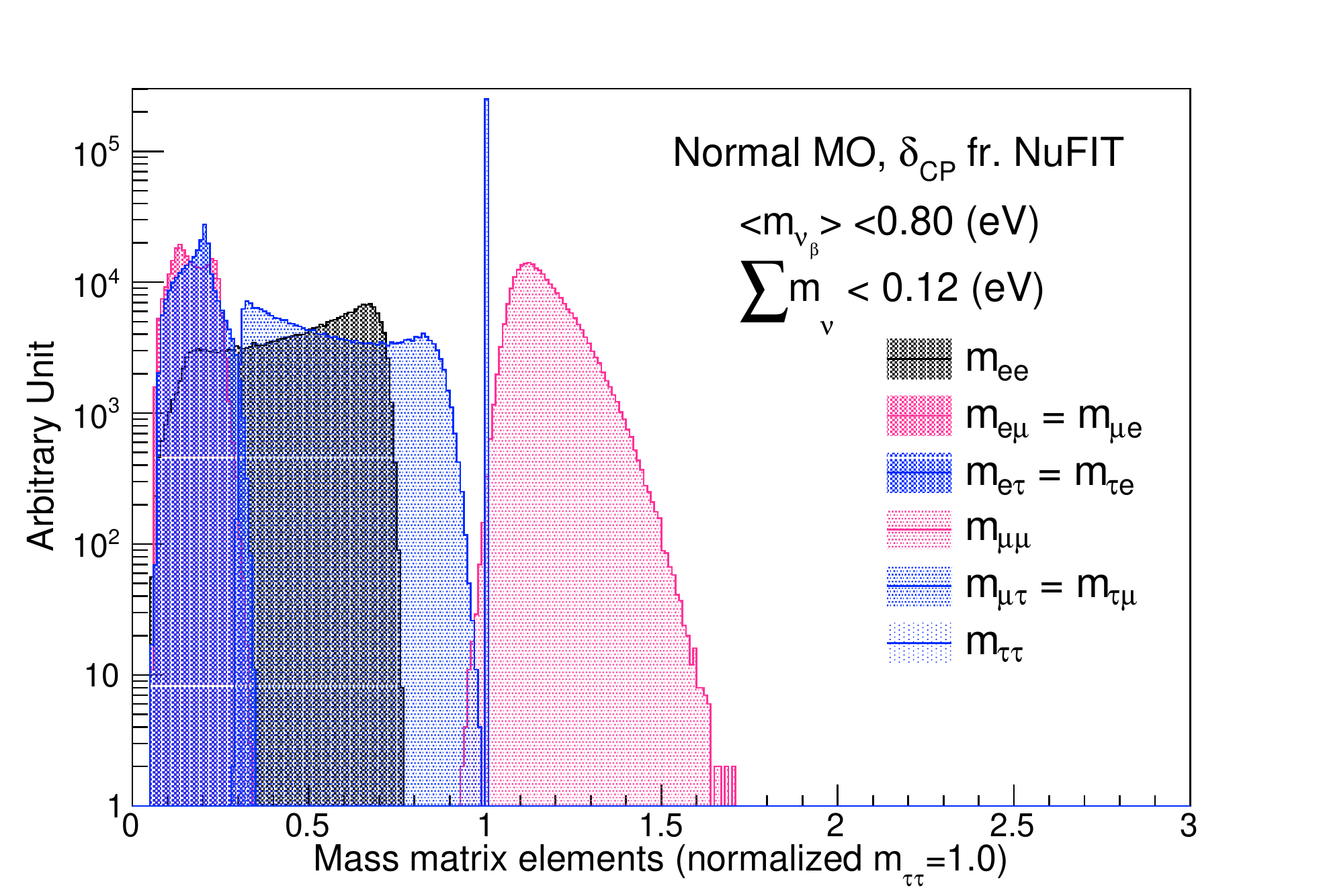}
\includegraphics[width=0.47\textwidth]{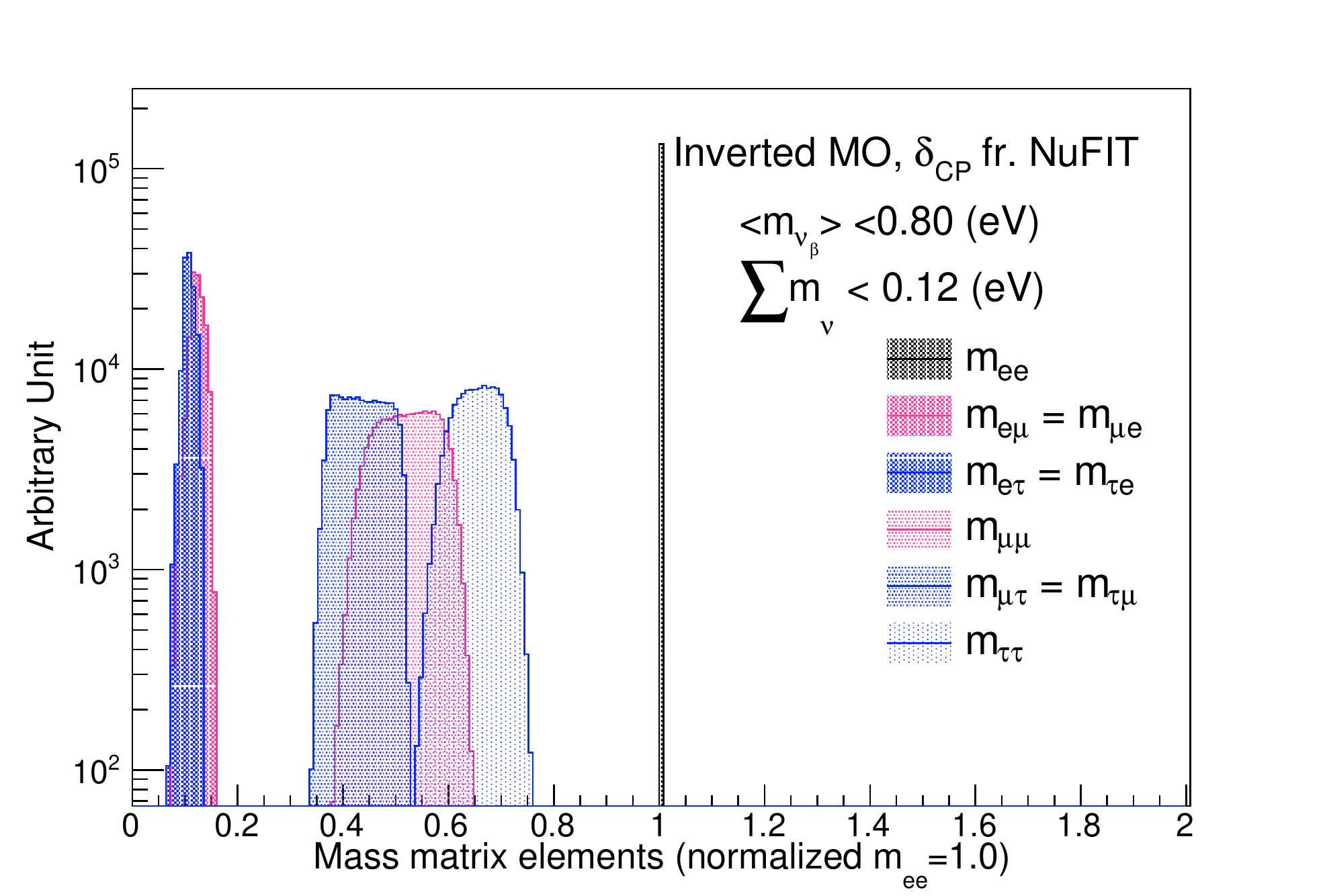}
\caption{Distributions of the elements' amplitude of the Majorana neutrino mass matrix $M_{\nu}^{Majorana}$ attained from the leptonic mixing matrix $U_{\text{PMNS}}$ via Eq.~(\ref{eq:majoranamassvspmns}). The present constraint of $\sum_{N_{\nu}} m_{\nu_i}<0.12\ \text{eV}$ from cosmology and the effective neutrino mass $\langle m_{\nu_{\beta}}\rangle<0.8 \ \text{eV}$ from the KATRIN experiment of the beta decay is used to limit the allowed phase space of the parameters. All elements are normalized to the $m_{\tau\tau}$ amplitude, i.e., $m_{\tau\tau}\equiv 1$ in the case of \emph{normal} MO and $m_{ee}$ amplitude, i.e., $m_{ee}\equiv 1$ in the case of \emph{inverted} MO. The left is with \emph{normal} neutrino MO and the right is with \emph{inverted} neutrino MO. The $\delta_{CP}$ constraint is derived from the global analysis~\cite{nufit5}. \label{fig:massmatrixcurrent}}
\end{figure}

As pointed out in Ref.~\cite{Mohapatra:1980yp}, it is model-dependent to work out the neutrino mass matrix from what we have actually observed, i.e. the leptonic mixing matrix. There is one case where we can have a direct connection between the PMNS leptonic mixing matrix and the neutrino mass matrix: when the neutrinos are Majorana particle and we assume to work in a weak-state basics, where the charged lepton mass matrix is diagonal. The scenario is in fact exciting since the smallness of neutrino mass can then be explained naturally by the seesaw mechanism~\cite{GellMann:1980vs,Yanagida:1979as,Mohapatra:1980yp} which is summarized in~\cite{Mohapatra:1999em}. In this case, the Majorana mass matrix $M_{\nu}^{Majorana}$ is related to the leptonic mixing matrix $U_{\text{PMNS}}$ via
\begin{align}\label{eq:majoranamassvspmns}
M_{\nu}^{\text{Majorana}} \equiv  \begin{pmatrix}
   m_{ee} & m_{e\mu} & m_{e\tau}\\
  m_{\mu e} & m_{\mu \mu} & m_{\mu \tau}\\
    m_{\tau e} & m_{\tau \mu} & m_{\tau \tau}
    \end{pmatrix} =  U_{\text{PMNS}}^{*}Diag(m_1,m_2,m_3)U_{\text{PMNS}}^{\dagger}
\end{align}
Apparently $M_{\nu}^{Majorana}$ is a symmetric $3\times3$ matrix. Fig.~\ref{fig:massmatrixcurrent} shows the distribution of the neutrino mass elements' amplitude with current limination on the leptonic mixing and the constraint from cosmology for the sum of neutrino mass, $\sum_{N_{\nu}} m_{\nu_i}<0.12\ \text{eV}$, and effective electron neutrino mass, $\langle m_{\nu_{\beta}}\rangle<0.8 \ \text{eV}$, from KATRIN experiment with the beta decay. The neutrino mass spectra are calculated for two cases: \emph{normal} and \emph{inverted} neutrino MO. The constraint of $\delta_{CP}$ is taken from the global analysis~\cite{nufit5}.  
\noindent One can observe from Fig.~\ref{fig:massmatrixcurrent} that for the scenario where the neutrino mass ordering is normal, the texture of the matrix is unclear. All matrix elements are at the $\mathcal{O}(1)$ level.  If the neutrino mass ordering is inverted and if we take care of the sub-$\mathcal{O}(1)$ level, the Majorana neutrino mass displays some discrete pattern like,
\begin{align}\label{eq:matrixmajoranacurrent}
  \left|M_{\nu, \text{Normal MO}}^{\text{Majorana}}\right|=
       \begin{pmatrix}
   0.47\pm0.18 & 0.17\pm0.05 & 0.18\pm 0.05\\
  0.17\pm0.05 & 1.17\pm0.09 & 0.57\pm 0.18\\
    0.19\pm 0.05 & 0.57\pm 0.18 & 1.
    \end{pmatrix};\nonumber \\ 
      \left|M_{\nu, \text{Inverted MO}}^{\text{Majorana}}\right|=
       \begin{pmatrix}
   1 & 0.12\pm0.01 & 0.11\pm 0.01\\
  0.12\pm0.01 & 0.52\pm0.06 & 0.44\pm 0.04\\
    0.11\pm 0.01 & 0.44\pm 0.04 & 0.65\pm 0.04
    \end{pmatrix}
\end{align}
For both cases of the neutrino MO, the largest ratio among the matrix elements is within $\mathcal{O}(1)$. Although the estimation is with a constraint of \dcp\ from the global fit, it is found that the pattern in Eq.~(\ref{eq:matrixmajoranacurrent}) stays almost the same at different fixed values of \dcp.
%%%%%%%%%%%%%%%%%%%%%%%%%%%%%%%%%%%%%%%%%%%%%%%%%%%
%%%%%%%%%%%%%%%%%%%%%%%%%%%%%%%%%%%%%%%%%%%%%%%%%%%
\section{Prospect of neutrino mass ordering determination}\label{sec:future}
\begin{figure}[H]
\centering
\includegraphics[width=0.7\textwidth]{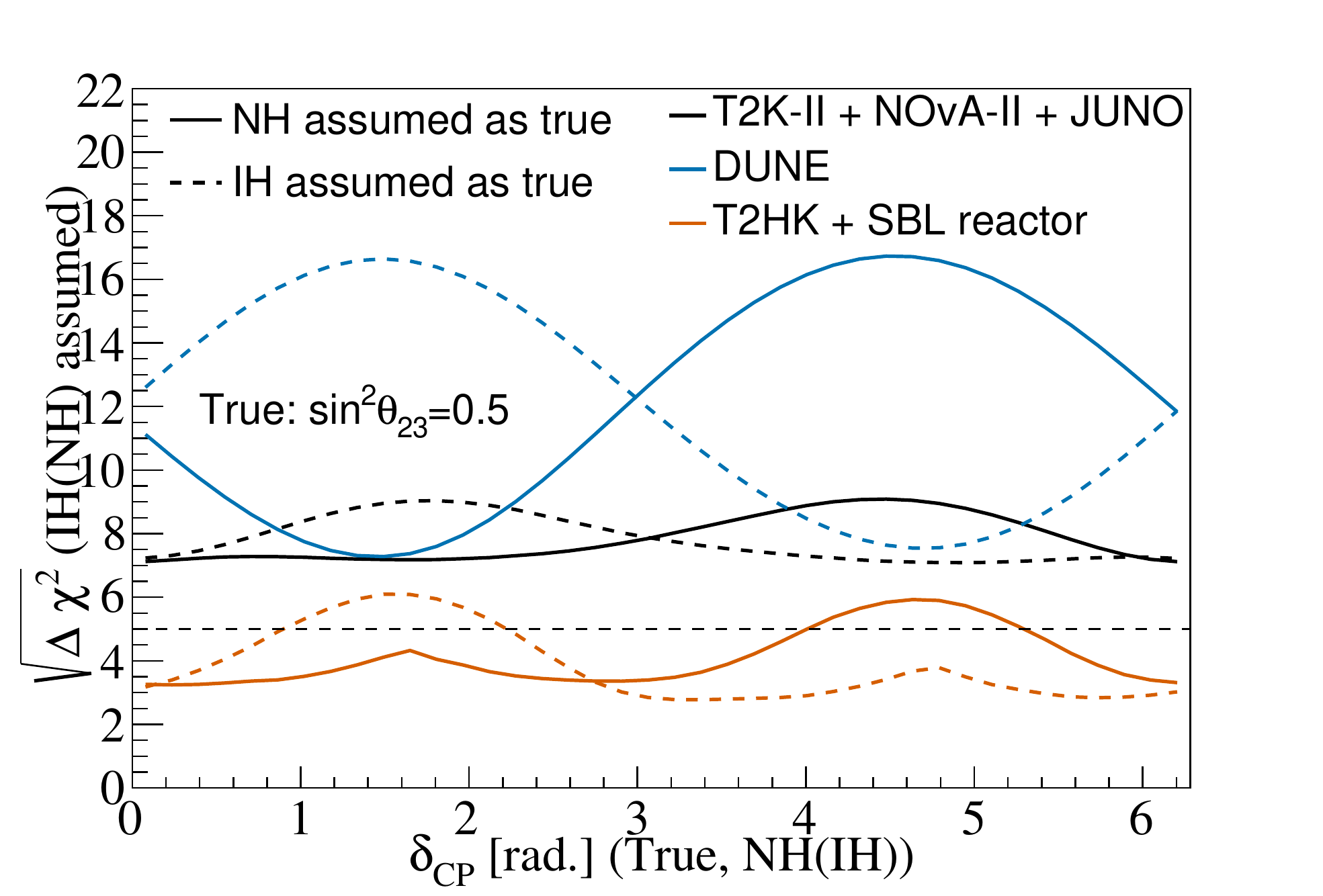}
\caption{Neutrino MO sensitivity as a function of the true \dcp\ for two possible MO hypotheses with a joint analysis of T2K-II, \nova-II, and JUNO, with DUNE, and with joint T2HK and short-baseline (SBL) reactor experiments.\label{fig:massoscconstraint}}
\end{figure}

On the absolute neutrino mass scale, KATRIN experiment aim to achieve an ultimate upper threshold of 0.2~eV~\cite{Aker:2021gma} for the effective electron neutrino mass $\langle m_{\nu_{\beta}}\rangle$. The Project 8~\cite{Esfahani:2017dmu} aims for reaching a sensitivity of $\langle m_{\nu_{\beta}}\rangle$ to 0.04~eV. The future of KamLAND2-Zen~\cite{Shirai:2017jyz} can reach to the level of 0.02~eV for effective Majorana mass 
$\langle m_{\beta\beta}^{0\nu}\rangle$. The next generation of the experiments aims to provide a 0.01~eV sensitivity of $\langle m_{\beta\beta}^{0\nu}\rangle$, with a maximum reach of 0.001~eV~\cite{Giuliani:2019uno}. The future cosmological constraint on $\sum_{N_{\nu}} m_{\nu_i}$ can reach 0.06~eV~\cite{Brinckmann:2018owf}, assuming the the lightest mass is zero and the total mass is about $\sqrt{\Delta m^2_{31}}+\sqrt{\Delta m^2_{21}}$. These future prospects are very encouraging, since as implied in Fig.~\ref{fig:massoscconstraint}, the inverted neutrino mass ordering will be significantly ruled out if the future constraints reveal one of the following scenarios:
\begin{itemize}
    \item Cosmology constraint will support $\sum_{N_{\nu}} m_{\nu_i}<0.1\text{ eV}$
    \item The effective neutrino mass from the beta-decay will be found to be $\langle m_{\nu_{\beta}}\rangle < 0.04\text{ eV}$ 
    \item The \nulessbb process will be discovered and the effective Majorana neutrino mass will be found to be $\langle m_{\beta\beta}^{0\nu}\rangle<0.09\text{ eV}$
\end{itemize}
Regarding the neutrino mass information attained from the neutrino oscillation experiment, the prospect of resolving the neutrino mass ordering in the upcoming decades is bright. JUNO~\cite{djurcic2015juno} can achieve solely 3-4$\sigma$ confidence level (C.L.) on the neutrino MO resolving without any dependence on the \dcp\ phase. Besides, the ongoing A-LBL experiments T2K and \nova\ can provide significant sensitivity to resolve the neutrino MO by extending their operational run to collect more valuable data until the beginning of the next generation of neutrino experiments---DUNE and Hyper-Kamiokande. A joint analysis study in Ref.~\cite{Cao:2020aig} concludes that the combination of JUNO, T2K-II, and \nova-II will determine distinctly the neutrino MO---more than 6$\sigma$ C.L. Besides, the DUNE experiment by itself can reach the answer conclusively---more than 6$\sigma$ C.L.---and Tokai to Hyper-Kamiokande (T2HK) experiment by itself can provide a significant resolving (3-6$\sigma$ C.L. depending on the true value of \dcp) to the neutrino mass ordering.
For this study, we use the DUNE detector simulation configuration ~\cite{DUNE:2016ymp}, our simulation to match the physics potential of the Hyper-Kamiokande~\cite{Hyper-Kamiokande:2018ofw}, and a study of combined on-going A-LBL neutrino experiments, T2K-II and \nova-II, and the R-MBL neutrino experiment---JUNO---in Ref.~\cite{Cao:2020aig}.

\begin{figure}[H]	
\centering
\includegraphics[width=0.47\textwidth]{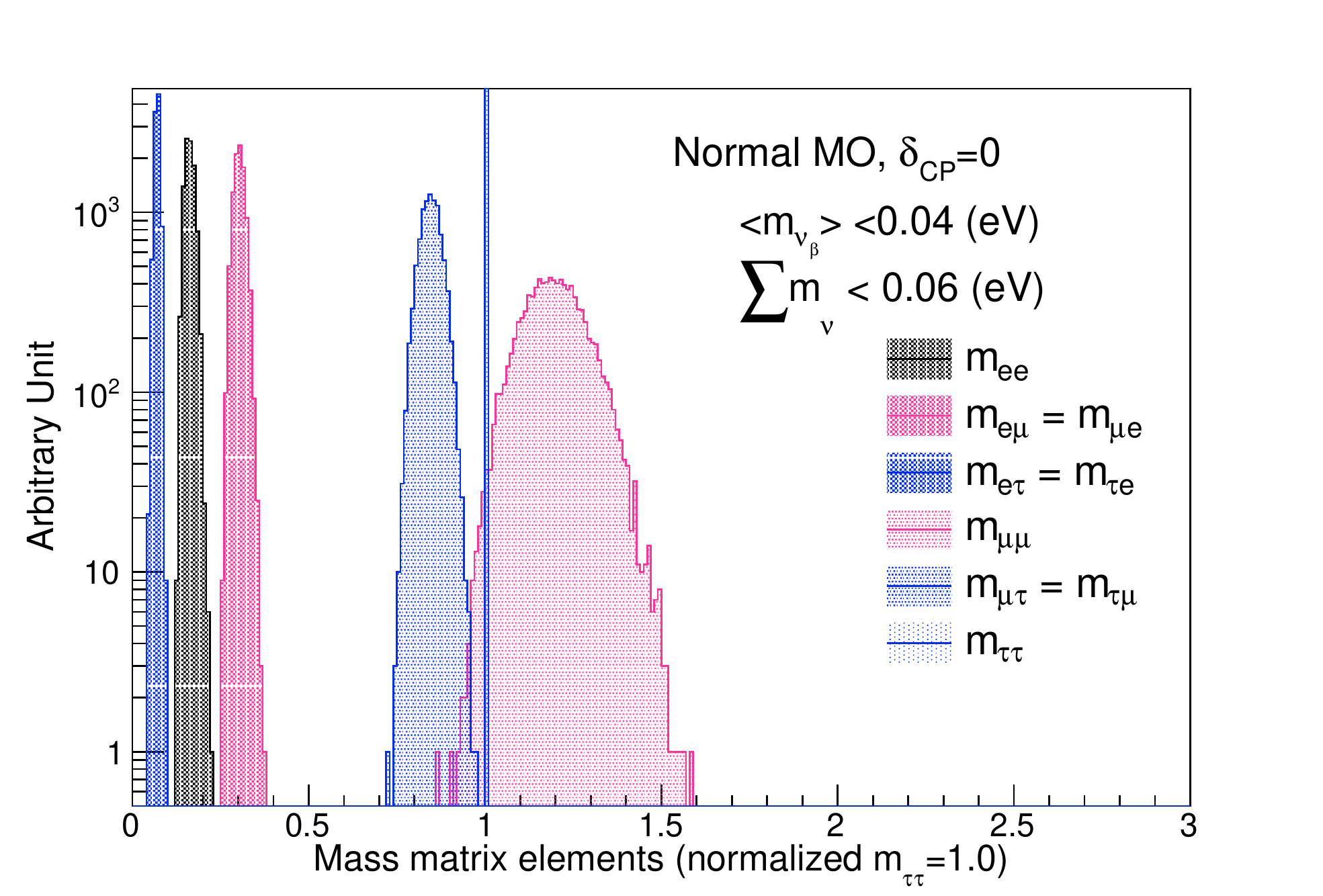}
\includegraphics[width=0.47\textwidth]{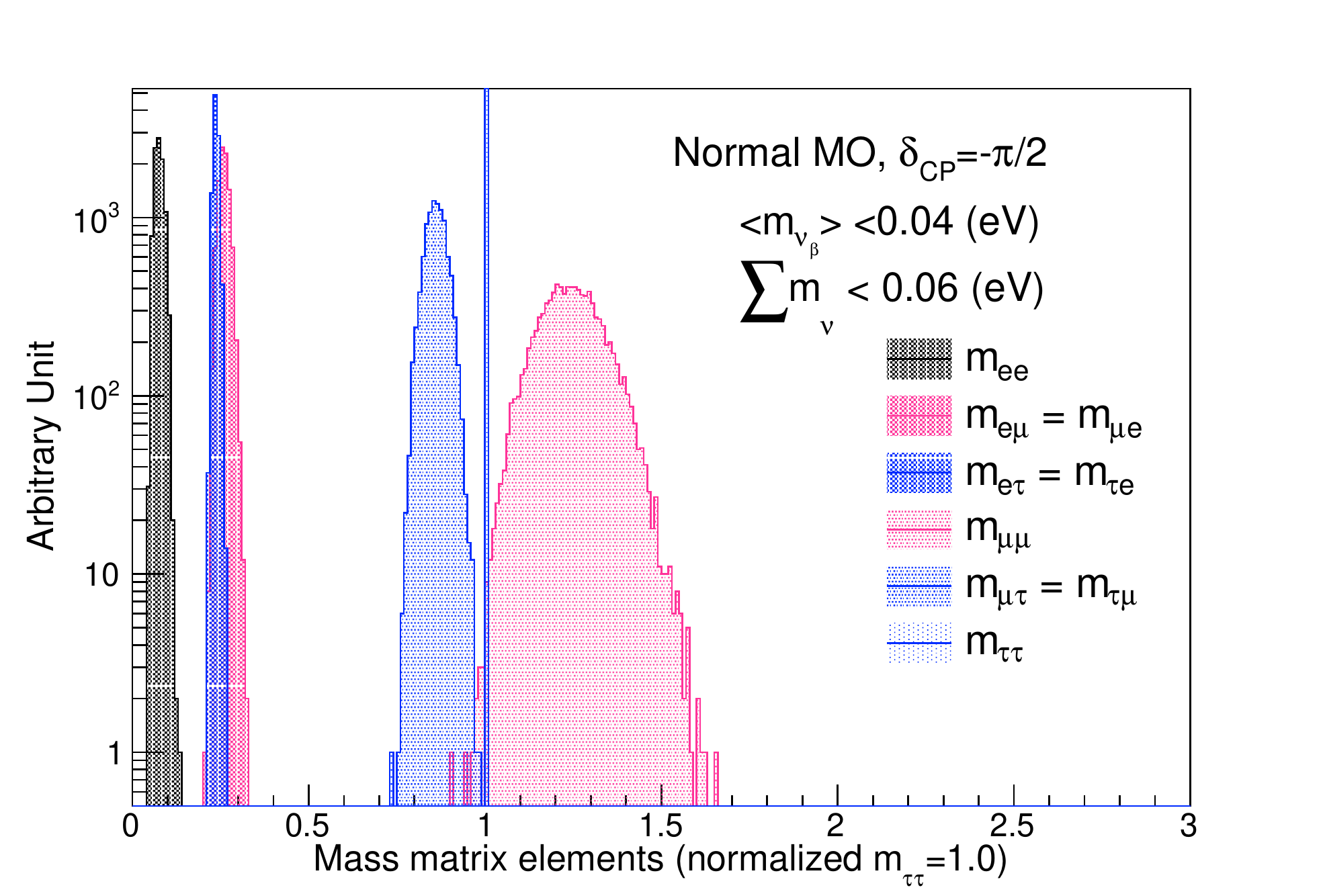}
\caption{Distributions of the elements' amplitude of the Majorana neutrino mass matrix $M_{\nu}^{Majorana}$ attained from the leptonic mixing matrix $U_{\text{PMNS}}$ via Eq.~(\ref{eq:majoranamassvspmns}). The future constraint of $\sum_{N_{\nu}} m_{\nu_i}<0.06\ \text{eV}$ from cosmology and the effective neutrino mass $\langle m_{\nu_{\beta}}\rangle<0.04 \ \text{eV}$ from the beta decay are used to limit the allowed phase space of the parameters. The $m_{\tau\tau}$ element is used to normalize all elements, i.e $m_{\tau\tau}\equiv 1$. The left has $\delta_{CP}=0$ and the right has $\delta_{CP}=-\pi/2$. Inverted MO will be ruled out in this hypothesis of the cosmological and beta-decay observations, and only normal MO will remain for investigation.\label{fig:massmatrixfuture}}
\end{figure}

\noindent As mentioned above, the future constraint on cosmology and beta decay can rule out the \emph{inverted} neutrino MO hypothesis if $\sum_{N_{\nu}} m_{\nu_i}<0.1\text{ eV}$ or $\langle m_{\nu_{\beta}}\rangle < 0.04\text{ eV}$. If this is the case, it is observed that the value of \dcp will have a strong impact on the observable pattern of the neutrino mass matrix elements. In Fig.~\ref{fig:massmatrixfuture}, we provide the patterns of the neutrino mass matrix elements for two extreme cases $\delta_{CP}=0$ and $\delta_{CP}=-\pi/2.$
All elements are in the $\mathcal{O}(1)$ level. If one cares about the sub-$\mathcal{O}(1)$ level, a notable difference in the pattern of the neutrino mass matrix can be observed with different truth values of \dcp. The rough estimation for the Majorana neutrino mass matrix pattern provides
\begin{align}\label{eq:matrixmajoranafuture}
     \left| M_{\nu, \text{Normal MO}}^{\text{Majorana}}\right|_{\delta_{CP}=0}&=
    \begin{pmatrix}
   0.16\pm 0.01 & 0.30\pm 0.02 & 0.07\pm 0.02\\
    0.30\pm 0.02 & 1.20\pm 0.09 & 0.85\pm 0.03\\
    0.07\pm 0.02 & 0.85\pm 0.03 & 1
    \end{pmatrix}; \nonumber \\
     \left| M_{\nu, \text{Normal MO}}^{\text{Majorana}}\right|_{\delta_{CP}=-\pi/2}&=
    \begin{pmatrix}
   0.08\pm 0.01 & 0.26\pm 0.02 & 0.24\pm 0.02\\
    0.26\pm 0.02 & 1.25\pm 0.09 & 0.86\pm 0.03\\
    0.24\pm 0.02 & 0.86\pm 0.03 & 1
    \end{pmatrix};
\end{align}
For both cases of the neutrino MO, the largest ratio among the matrix elements is within $\mathcal{O}(10)$.
\noindent If the underlying neutrino mass pattern is manifested by an unknown flavor symmetry, we find it interesting and encouraging to not only search for the CPV in the leptonic mixing matrix but also measure the $\delta_{CP}$ with high precision.

%%%%%%%%%%%%%%%%%%%%%%%%%%%%%%%%%%%%%%%%%%

%%%%%%%%%%%%%%%%%%%%%%%%%%%%%%%%%%%%%%%%%%
\section{Discussion}\label{sec:discuss}
 Froggatt-Nielsen mechanism~\cite{froggatt1979hierarchy} is among the earliest and most interesting attempts to explain the smallness and hierarchy of the charged fermion mass. By introducing an approximate $U(1)_{FN}$ to provide various charge to the fermion, the mass matrix for up-type quarks and down-type quarks~\cite{nir2016flavour} can be read as 
 \begin{align}\label{eq:matrixquark}
      Y_{u-type}=
    \begin{pmatrix}
   \epsilon_{\text{FN}}^4 & \epsilon_{\text{FN}}^{3} & \epsilon_{\text{FN}}^{2}\\
    \epsilon_{\text{FN}}^{3} & \epsilon_{\text{FN}}^{2} & \epsilon_{\text{FN}}\\
    \epsilon_{\text{FN}}^{2} & \epsilon_{\text{FN}} & 1
    \end{pmatrix}, \
Y_{d-type}=Y_{e}^{T}=
    \begin{pmatrix}
   \epsilon_{\text{FN}}^2 & \epsilon_{\text{FN}}^{2} & \epsilon_{\text{FN}}^{2}\\
    \epsilon_{\text{FN}} & \epsilon_{\text{FN}} & \epsilon_{\text{FN}}\\
    1 & 1 & 1
    \end{pmatrix},
\end{align}
where $\epsilon_{\text{FN}} \sim \sin^2 \theta_{\text{Cabbio}} \sim 0.05$. However, indirect observation of the neutrino mass matrix seems to not exhibit any of the above-mentioned patterns. No matter what neutrino MO can be, we find that $m_{\nu2}/m_{\nu3}$>0.17, which is much larger than the counterparts of the charged fermion, giving $m_{\mu}/m_{\tau}$ = 0.06, for up-type quarks $m_{\text{c-quark}}/m_{\text{t-quark}}=0.01$, down-type quarks $m_{\text{s-quark}}/m_{\text{b-quark}}=0.02$. Also, in a special case where we can obtain the neutrino mass matrix directly from the leptonic mixing matrix, the patterns displayed in Eq.~(\ref{eq:matrixmajoranacurrent}) and Eq.~(\ref{eq:matrixmajoranafuture}) fairly differentiate them from the charged fermion pattern in Eq.~(\ref{eq:matrixquark}). 

\begin{figure}[H]	
\centering
\includegraphics[width=0.7\textwidth]{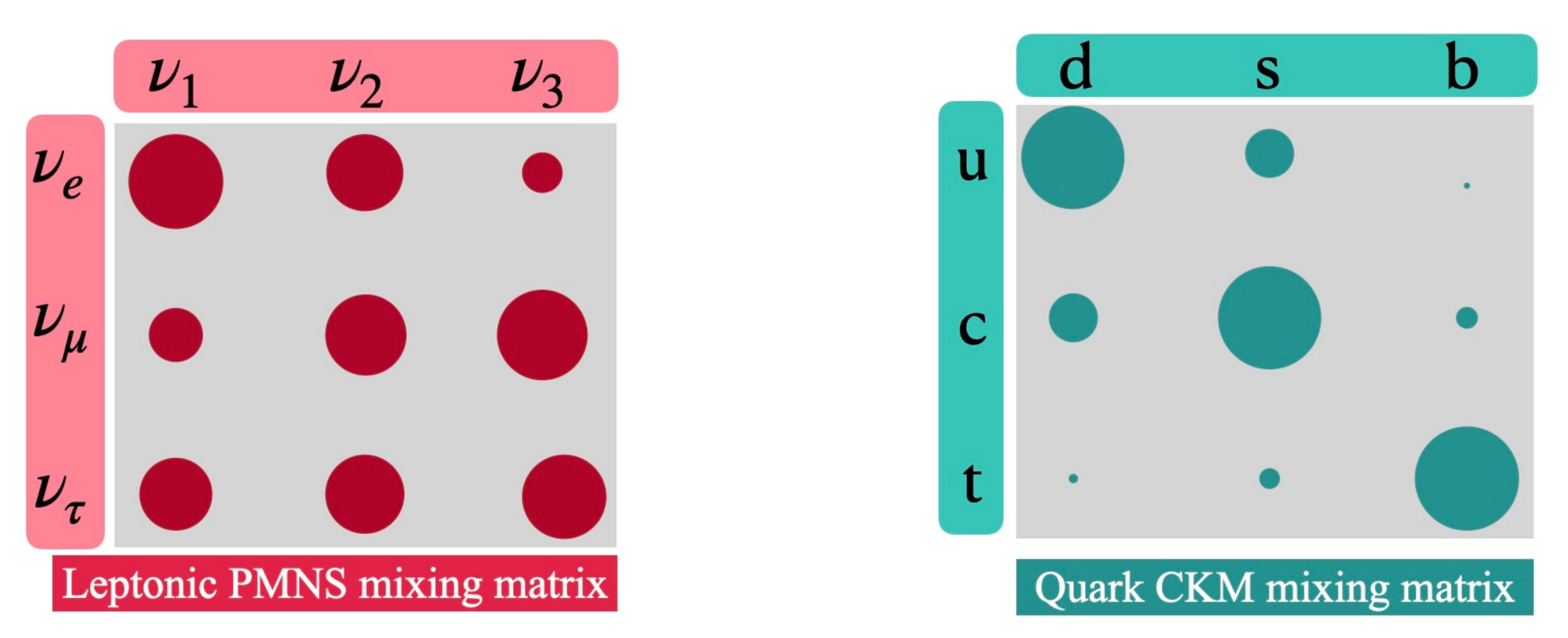}
\caption{The pictorial representation of the leptonic PMNS mixing matrix (left) vs. the quark CKM mixing matrix (right). The dot area is proportional to the matrix element's amplitude. The quark mixing parameters are taken from Ref.~\cite{Zyla:2020zbs}. The leptonic mixing parameters are the same as in Table~\ref{tab:nuoscpara}. \label{fig:nupmnsvsckm}}
\end{figure}

In addition, it is observed that the quark CKM mixing matrices and the leptonic PMNS mixing matrices is sizably different. While the former is close to the identity, almost all elements of the latter are sizeable, as shown in Fig.~\ref{fig:nupmnsvsckm}.  
%\begin{align*}
 %   \sin^2\theta_{12} = 0.22650 \pm 0.00048; \ \sin^2\theta_{23} = 0.04053^{+0.00083}_{-0.00061}; \ 
  %  \sin62\theta_{13} = 0.00361^{+0.00011}_{-0.00009}; \
   % \delta_{\text{CP}} = 1.196^{+0.045}_{-0.043}
%\end{align*}

\noindent One interesting idea to explain the leptonic PMNS mixing matrix is the manifestation of the neutrino mass anarchy proposed in Ref.~\cite{Hall:1999sn,Haba:2000be,deGouvea:2003xe}. Central of the approach is that there is no special texture in the neutrino mass matrix, meaning that all elements of the matrix are at the same order of the magnitude and can be drawn randomly. The present data as well as the reconstructed neutrino mass matrices under particular assumption in Eq.~(\ref{eq:matrixmajoranacurrent}) show good agreement with the anarchy hypothesis. Furthermore, even if the future constraint on the mass observables leads to the neutrino mass pattern displayed in Eq.~(\ref{eq:matrixmajoranafuture}), we cannot falsify the anarchy hypothesis. If this hypothesis is true, the anarchy-like neutrino mass spectrum distinguished from the hierarchy of the charged fermion mass can be interpreted as the difference in the mass generation mechanism. Neutrino mass may not come from the Yukawa interaction with the SM Higgs boson but from others, such as the seesaw mechanism. 

For the two undetermined parameters in the neutrino oscillations---neutrino MO and value of \dcp---one may need to reconsider the anarchy approach if \dcp is found to be close to zero, i.e no significant CP violation is found since there is no reason for CP symmetry to be conserved in this approach. Whether neutrino MO is \emph{normal} or \emph{inverted} cannot help us select the hierarchy or anarchy approach as Nature's favor unless some clearer texture of the neutrino mass matrix is unraveled in the future.

%%%%%%%%%%%%%%%%%%%%%%%%%%%%%%%%%%%%%%%%%%
\section{Conclusions}
The determination of the neutrino mass spectrum is critical not only for completing the landscape of leptonic mixing but also for providing an important input for extending the Standard Model. We have reviewed and made some observations about the neutrino mass spectrum using the current the neutrino oscillation data, cosmology, and beta-decay constraints. At the moment, it is not clear whether neutrino mass ordering is \emph{normal} or \emph{inverted}. For illustration, we reconstruct the neutrino mass matrix under a specific assumption about neutrinos' characteristics and find that it consistent with the neutrino mass anarchy hypothesis. We also calculate the sensitivity of the on-going and future neutrino experiments to determine the neutrino mass ordering. The conclusion is that this unknown will be distinctly elucidated by an extension of the current neutrino programs or by the next generation of the neutrino experiments. Furthermore, our research contributes to future programs that precisely measure the \dcp\ phase, allowing us to better establish the neutrino mass spectrum and shed light on the underlying neutrino mass generation model. 
\acknowledgments{S. Cao would like to thank Neutrino Group, IPNS, KEK for their hospitality. This work is partially supported by the VAST-JSPS joint program under grant No. QTJP01.02/20-22. T. V. Ngoc was funded by Vingroup JSC and supported by the Master, PhD Scholarship Programme of Vingroup Innovation Foundation (VINIF), Institute of Big Data, code VINIF.2021.TS.069}

\bibliography{ref}% Produces the

\begin{thebibliography}{41}%
\makeatletter
\providecommand \@ifxundefined [1]{%
 \@ifx{#1\undefined}
}%
\providecommand \@ifnum [1]{%
 \ifnum #1\expandafter \@firstoftwo
 \else \expandafter \@secondoftwo
 \fi
}%
\providecommand \@ifx [1]{%
 \ifx #1\expandafter \@firstoftwo
 \else \expandafter \@secondoftwo
 \fi
}%
\providecommand \natexlab [1]{#1}%
\providecommand \enquote  [1]{``#1''}%
\providecommand \bibnamefont  [1]{#1}%
\providecommand \bibfnamefont [1]{#1}%
\providecommand \citenamefont [1]{#1}%
\providecommand \href@noop [0]{\@secondoftwo}%
\providecommand \href [0]{\begingroup \@sanitize@url \@href}%
\providecommand \@href[1]{\@@startlink{#1}\@@href}%
\providecommand \@@href[1]{\endgroup#1\@@endlink}%
\providecommand \@sanitize@url [0]{\catcode `\\12\catcode `\$12\catcode
  `\&12\catcode `\#12\catcode `\^12\catcode `\_12\catcode `\%12\relax}%
\providecommand \@@startlink[1]{}%
\providecommand \@@endlink[0]{}%
\providecommand \url  [0]{\begingroup\@sanitize@url \@url }%
\providecommand \@url [1]{\endgroup\@href {#1}{\urlprefix }}%
\providecommand \urlprefix  [0]{URL }%
\providecommand \Eprint [0]{\href }%
\providecommand \doibase [0]{http://dx.doi.org/}%
\providecommand \selectlanguage [0]{\@gobble}%
\providecommand \bibinfo  [0]{\@secondoftwo}%
\providecommand \bibfield  [0]{\@secondoftwo}%
\providecommand \translation [1]{[#1]}%
\providecommand \BibitemOpen [0]{}%
\providecommand \bibitemStop [0]{}%
\providecommand \bibitemNoStop [0]{.\EOS\space}%
\providecommand \EOS [0]{\spacefactor3000\relax}%
\providecommand \BibitemShut  [1]{\csname bibitem#1\endcsname}%
\let\auto@bib@innerbib\@empty
%</preamble>
\bibitem [{\citenamefont {Pauli}(1978)}]{Pauli:1930pc}%
  \BibitemOpen
  \bibfield  {author} {\bibinfo {author} {\bibfnamefont {W.}~\bibnamefont
  {Pauli}},\ }\href@noop {} {\bibfield  {journal} {\bibinfo  {journal} {Phys.
  Today}\ }\textbf {\bibinfo {volume} {31N9}},\ \bibinfo {pages} {27} (\bibinfo
  {year} {1978})}\BibitemShut {NoStop}%
\bibitem [{\citenamefont {Reines}\ and\ \citenamefont
  {Cowan}(1956)}]{Reines:1956rs}%
  \BibitemOpen
  \bibfield  {author} {\bibinfo {author} {\bibfnamefont {F.}~\bibnamefont
  {Reines}}\ and\ \bibinfo {author} {\bibfnamefont {C.~L.}\ \bibnamefont
  {Cowan}},\ }\href {\doibase 10.1038/178446a0} {\bibfield  {journal} {\bibinfo
   {journal} {Nature}\ }\textbf {\bibinfo {volume} {178}},\ \bibinfo {pages}
  {446} (\bibinfo {year} {1956})}\BibitemShut {NoStop}%
\bibitem [{\citenamefont {Goldhaber}\ \emph {et~al.}(1958)\citenamefont
  {Goldhaber}, \citenamefont {Grodzins},\ and\ \citenamefont
  {Sunyar}}]{Goldhaber:1958nb}%
  \BibitemOpen
  \bibfield  {author} {\bibinfo {author} {\bibfnamefont {M.}~\bibnamefont
  {Goldhaber}}, \bibinfo {author} {\bibfnamefont {L.}~\bibnamefont {Grodzins}},
  \ and\ \bibinfo {author} {\bibfnamefont {A.~W.}\ \bibnamefont {Sunyar}},\
  }\href {\doibase 10.1103/PhysRev.109.1015} {\bibfield  {journal} {\bibinfo
  {journal} {Phys. Rev.}\ }\textbf {\bibinfo {volume} {109}},\ \bibinfo {pages}
  {1015} (\bibinfo {year} {1958})}\BibitemShut {NoStop}%
\bibitem [{\citenamefont {Zyla}\ \emph {et~al.}(2020)\citenamefont {Zyla} \emph
  {et~al.}}]{Zyla:2020zbs}%
  \BibitemOpen
  \bibfield  {author} {\bibinfo {author} {\bibfnamefont {P.~A.}\ \bibnamefont
  {Zyla}} \emph {et~al.} (\bibinfo {collaboration} {Particle Data Group}),\
  }\href {\doibase 10.1093/ptep/ptaa104} {\bibfield  {journal} {\bibinfo
  {journal} {PTEP}\ }\textbf {\bibinfo {volume} {2020}},\ \bibinfo {pages}
  {083C01} (\bibinfo {year} {2020})}\BibitemShut {NoStop}%
\bibitem [{\citenamefont {Esteban}\ \emph {et~al.}(2019)\citenamefont
  {Esteban}, \citenamefont {Gonzalez-Garcia}, \citenamefont
  {Hernandez-Cabezudo}, \citenamefont {Maltoni},\ and\ \citenamefont
  {Schwetz}}]{esteban2019global}%
  \BibitemOpen
  \bibfield  {author} {\bibinfo {author} {\bibfnamefont {I.}~\bibnamefont
  {Esteban}}, \bibinfo {author} {\bibfnamefont {M.}~\bibnamefont
  {Gonzalez-Garcia}}, \bibinfo {author} {\bibfnamefont {A.}~\bibnamefont
  {Hernandez-Cabezudo}}, \bibinfo {author} {\bibfnamefont {M.}~\bibnamefont
  {Maltoni}}, \ and\ \bibinfo {author} {\bibfnamefont {T.}~\bibnamefont
  {Schwetz}},\ }\href {\doibase 10.1007/JHEP01(2019)106} {\bibfield  {journal}
  {\bibinfo  {journal} {JHEP}\ }\textbf {\bibinfo {volume} {01}},\ \bibinfo
  {pages} {106} (\bibinfo {year} {2019})},\ \Eprint
  {http://arxiv.org/abs/1811.05487} {arXiv:1811.05487 [hep-ph]} \BibitemShut
  {NoStop}%
\bibitem [{\citenamefont {Esteban}\ \emph
  {et~al.}(2020{\natexlab{a}})\citenamefont {Esteban}, \citenamefont
  {Gonzalez-Garcia}, \citenamefont {Maltoni}, \citenamefont {Schwetz},\ and\
  \citenamefont {Zhou}}]{Esteban:2020cvm}%
  \BibitemOpen
  \bibfield  {author} {\bibinfo {author} {\bibfnamefont {I.}~\bibnamefont
  {Esteban}}, \bibinfo {author} {\bibfnamefont {M.}~\bibnamefont
  {Gonzalez-Garcia}}, \bibinfo {author} {\bibfnamefont {M.}~\bibnamefont
  {Maltoni}}, \bibinfo {author} {\bibfnamefont {T.}~\bibnamefont {Schwetz}}, \
  and\ \bibinfo {author} {\bibfnamefont {A.}~\bibnamefont {Zhou}},\ }\href
  {\doibase 10.1007/JHEP09(2020)178} {\bibfield  {journal} {\bibinfo  {journal}
  {JHEP}\ }\textbf {\bibinfo {volume} {09}},\ \bibinfo {pages} {178} (\bibinfo
  {year} {2020}{\natexlab{a}})},\ \Eprint {http://arxiv.org/abs/2007.14792}
  {arXiv:2007.14792 [hep-ph]} \BibitemShut {NoStop}%
\bibitem [{\citenamefont {Esteban}\ \emph
  {et~al.}(2020{\natexlab{b}})\citenamefont {Esteban} \emph {et~al.}}]{nufit5}%
  \BibitemOpen
  \bibfield  {author} {\bibinfo {author} {\bibfnamefont {I.}~\bibnamefont
  {Esteban}} \emph {et~al.},\ }\href@noop {} {\enquote {\bibinfo {title} {Nufit
  5.0 (2020), www.nu-fit.org},}\ }\bibinfo {howpublished}
  {\url{http://www.nu-fit.org/?q=node/228}} (\bibinfo {year}
  {2020}{\natexlab{b}})\BibitemShut {NoStop}%
\bibitem [{\citenamefont {Wolfenstein}(1978)}]{Wolfenstein:1977ue}%
  \BibitemOpen
  \bibfield  {author} {\bibinfo {author} {\bibfnamefont {L.}~\bibnamefont
  {Wolfenstein}},\ }\href {\doibase 10.1103/PhysRevD.17.2369} {\bibfield
  {journal} {\bibinfo  {journal} {Phys. Rev. D}\ }\textbf {\bibinfo {volume}
  {17}},\ \bibinfo {pages} {2369} (\bibinfo {year} {1978})}\BibitemShut
  {NoStop}%
\bibitem [{\citenamefont {Mikheyev}\ and\ \citenamefont
  {Smirnov}(1985)}]{Mikheyev:1985zog}%
  \BibitemOpen
  \bibfield  {author} {\bibinfo {author} {\bibfnamefont {S.~P.}\ \bibnamefont
  {Mikheyev}}\ and\ \bibinfo {author} {\bibfnamefont {A.~Y.}\ \bibnamefont
  {Smirnov}},\ }\href@noop {} {\bibfield  {journal} {\bibinfo  {journal} {Sov.
  J. Nucl. Phys.}\ }\textbf {\bibinfo {volume} {42}},\ \bibinfo {pages} {913}
  (\bibinfo {year} {1985})}\BibitemShut {NoStop}%
\bibitem [{\citenamefont {Mikheev}\ and\ \citenamefont
  {Smirnov}(1986)}]{Mikheev:1986wj}%
  \BibitemOpen
  \bibfield  {author} {\bibinfo {author} {\bibfnamefont {S.~P.}\ \bibnamefont
  {Mikheev}}\ and\ \bibinfo {author} {\bibfnamefont {A.~Y.}\ \bibnamefont
  {Smirnov}},\ }\href {\doibase 10.1007/BF02508049} {\bibfield  {journal}
  {\bibinfo  {journal} {Nuovo Cim. C}\ }\textbf {\bibinfo {volume} {9}},\
  \bibinfo {pages} {17} (\bibinfo {year} {1986})}\BibitemShut {NoStop}%
\bibitem [{\citenamefont {Bellini}\ \emph {et~al.}(2012)\citenamefont {Bellini}
  \emph {et~al.}}]{Borexino:2011ufb}%
  \BibitemOpen
  \bibfield  {author} {\bibinfo {author} {\bibfnamefont {G.}~\bibnamefont
  {Bellini}} \emph {et~al.} (\bibinfo {collaboration} {Borexino}),\ }\href
  {\doibase 10.1103/PhysRevLett.108.051302} {\bibfield  {journal} {\bibinfo
  {journal} {Phys. Rev. Lett.}\ }\textbf {\bibinfo {volume} {108}},\ \bibinfo
  {pages} {051302} (\bibinfo {year} {2012})},\ \Eprint
  {http://arxiv.org/abs/1110.3230} {arXiv:1110.3230 [hep-ex]} \BibitemShut
  {NoStop}%
\bibitem [{\citenamefont {Dunne}(2020)}]{patrick_dunne_2020_4154355}%
  \BibitemOpen
  \bibfield  {author} {\bibinfo {author} {\bibfnamefont {P.}~\bibnamefont
  {Dunne}},\ }\href {\doibase 10.5281/zenodo.4154355} {\enquote {\bibinfo
  {title} {Latest neutrino oscillation results from t2k},}\ } (\bibinfo {year}
  {2020})\BibitemShut {NoStop}%
\bibitem [{\citenamefont {Himmel}(2020)}]{alex_himmel_2020_4142045}%
  \BibitemOpen
  \bibfield  {author} {\bibinfo {author} {\bibfnamefont {A.}~\bibnamefont
  {Himmel}},\ }\href {\doibase 10.5281/zenodo.4142045} {\enquote {\bibinfo
  {title} {New oscillation results from the nova experiment},}\ } (\bibinfo
  {year} {2020})\BibitemShut {NoStop}%
\bibitem [{\citenamefont {Nakajima}(2020)}]{yasuhiro_nakajima_2020_4134680}%
  \BibitemOpen
  \bibfield  {author} {\bibinfo {author} {\bibfnamefont {Y.}~\bibnamefont
  {Nakajima}},\ }\href {\doibase 10.5281/zenodo.4134680} {\enquote {\bibinfo
  {title} {{Recent results and future prospects from Super-Kamiokande}},}\ }
  (\bibinfo {year} {2020})\BibitemShut {NoStop}%
\bibitem [{\citenamefont {Kelly}\ \emph {et~al.}(2021)\citenamefont {Kelly},
  \citenamefont {Machado}, \citenamefont {Parke}, \citenamefont
  {Perez-Gonzalez},\ and\ \citenamefont {Funchal}}]{Kelly:2020fkv}%
  \BibitemOpen
  \bibfield  {author} {\bibinfo {author} {\bibfnamefont {K.~J.}\ \bibnamefont
  {Kelly}}, \bibinfo {author} {\bibfnamefont {P.~A.~N.}\ \bibnamefont
  {Machado}}, \bibinfo {author} {\bibfnamefont {S.~J.}\ \bibnamefont {Parke}},
  \bibinfo {author} {\bibfnamefont {Y.~F.}\ \bibnamefont {Perez-Gonzalez}}, \
  and\ \bibinfo {author} {\bibfnamefont {R.~Z.}\ \bibnamefont {Funchal}},\
  }\href {\doibase 10.1103/PhysRevD.103.013004} {\bibfield  {journal} {\bibinfo
   {journal} {Phys. Rev. D}\ }\textbf {\bibinfo {volume} {103}},\ \bibinfo
  {pages} {013004} (\bibinfo {year} {2021})},\ \Eprint
  {http://arxiv.org/abs/2007.08526} {arXiv:2007.08526 [hep-ph]} \BibitemShut
  {NoStop}%
\bibitem [{\citenamefont {Suekane}(2015)}]{Suekane:2015yta}%
  \BibitemOpen
  \bibfield  {author} {\bibinfo {author} {\bibfnamefont {F.}~\bibnamefont
  {Suekane}},\ }\href {\doibase 10.1007/978-4-431-55462-2} {\emph {\bibinfo
  {title} {{Neutrino Oscillations}: {A Practical Guide to Basics and
  Applications}}}},\ Vol.\ \bibinfo {volume} {898}\ (\bibinfo {year}
  {2015})\BibitemShut {NoStop}%
\bibitem [{\citenamefont {Djurcic}\ \emph {et~al.}(2015)\citenamefont {Djurcic}
  \emph {et~al.}}]{djurcic2015juno}%
  \BibitemOpen
  \bibfield  {author} {\bibinfo {author} {\bibfnamefont {Z.}~\bibnamefont
  {Djurcic}} \emph {et~al.} (\bibinfo {collaboration} {JUNO Collaboration}),\
  }\href@noop {} {\  (\bibinfo {year} {2015})},\ \Eprint
  {http://arxiv.org/abs/1508.07166} {arXiv:1508.07166 [physics.ins-det]}
  \BibitemShut {NoStop}%
\bibitem [{\citenamefont {Petcov}\ and\ \citenamefont
  {Piai}(2002)}]{petcov2002lma}%
  \BibitemOpen
  \bibfield  {author} {\bibinfo {author} {\bibfnamefont {S.}~\bibnamefont
  {Petcov}}\ and\ \bibinfo {author} {\bibfnamefont {M.}~\bibnamefont {Piai}},\
  }\href@noop {} {\bibfield  {journal} {\bibinfo  {journal} {Physics Letters
  B}\ }\textbf {\bibinfo {volume} {533}},\ \bibinfo {pages} {94} (\bibinfo
  {year} {2002})}\BibitemShut {NoStop}%
\bibitem [{\citenamefont {Abusleme}\ \emph {et~al.}(2021)\citenamefont
  {Abusleme} \emph {et~al.}}]{Abusleme:2020lur}%
  \BibitemOpen
  \bibfield  {author} {\bibinfo {author} {\bibfnamefont {A.}~\bibnamefont
  {Abusleme}} \emph {et~al.} (\bibinfo {collaboration} {JUNO Collaboration}),\
  }\href {\doibase 10.1007/JHEP03(2021)004} {\bibfield  {journal} {\bibinfo
  {journal} {J. High Energy Phys.}\ }\textbf {\bibinfo {volume} {03}},\
  \bibinfo {pages} {004} (\bibinfo {year} {2021})},\ \Eprint
  {http://arxiv.org/abs/2011.06405} {arXiv:2011.06405 [physics.ins-det]}
  \BibitemShut {NoStop}%
\bibitem [{\citenamefont {Cao}\ \emph {et~al.}(2021)\citenamefont {Cao},
  \citenamefont {Nath}, \citenamefont {Ngoc}, \citenamefont {Francis},
  \citenamefont {T.},\ and\ \citenamefont {Quyen}}]{Cao:2020aig}%
  \BibitemOpen
  \bibfield  {author} {\bibinfo {author} {\bibfnamefont {S.}~\bibnamefont
  {Cao}}, \bibinfo {author} {\bibfnamefont {A.}~\bibnamefont {Nath}}, \bibinfo
  {author} {\bibfnamefont {T.~V.}\ \bibnamefont {Ngoc}}, \bibinfo {author}
  {\bibfnamefont {N.~K.}\ \bibnamefont {Francis}}, \bibinfo {author}
  {\bibfnamefont {H.~V.~N.}\ \bibnamefont {T.}}, \ and\ \bibinfo {author}
  {\bibfnamefont {P.~T.}\ \bibnamefont {Quyen}},\ }\href {\doibase
  10.1103/PhysRevD.103.112010} {\bibfield  {journal} {\bibinfo  {journal}
  {Phys. Rev. D}\ }\textbf {\bibinfo {volume} {103}},\ \bibinfo {pages}
  {112010} (\bibinfo {year} {2021})},\ \Eprint
  {http://arxiv.org/abs/2009.08585} {arXiv:2009.08585 [hep-ph]} \BibitemShut
  {NoStop}%
\bibitem [{\citenamefont {Curran}\ \emph {et~al.}(1948)\citenamefont {Curran},
  \citenamefont {Angus},\ and\ \citenamefont {Cockcroft}}]{curran1948beta}%
  \BibitemOpen
  \bibfield  {author} {\bibinfo {author} {\bibfnamefont {S.}~\bibnamefont
  {Curran}}, \bibinfo {author} {\bibfnamefont {J.}~\bibnamefont {Angus}}, \
  and\ \bibinfo {author} {\bibfnamefont {A.}~\bibnamefont {Cockcroft}},\
  }\href@noop {} {\bibfield  {journal} {\bibinfo  {journal} {Nature}\ }\textbf
  {\bibinfo {volume} {162}},\ \bibinfo {pages} {302} (\bibinfo {year}
  {1948})}\BibitemShut {NoStop}%
\bibitem [{\citenamefont {Aker}\ \emph {et~al.}(2021)\citenamefont {Aker} \emph
  {et~al.}}]{Aker:2021gma}%
  \BibitemOpen
  \bibfield  {author} {\bibinfo {author} {\bibfnamefont {M.}~\bibnamefont
  {Aker}} \emph {et~al.},\ }\href@noop {} {\  (\bibinfo {year} {2021})},\
  \Eprint {http://arxiv.org/abs/2105.08533} {arXiv:2105.08533 [hep-ex]}
  \BibitemShut {NoStop}%
\bibitem [{\citenamefont {de~Gouvea}\ and\ \citenamefont
  {Jenkins}(2008)}]{deGouvea:2008nm}%
  \BibitemOpen
  \bibfield  {author} {\bibinfo {author} {\bibfnamefont {A.}~\bibnamefont
  {de~Gouvea}}\ and\ \bibinfo {author} {\bibfnamefont {J.}~\bibnamefont
  {Jenkins}},\ }\href {\doibase 10.1103/PhysRevD.78.053003} {\bibfield
  {journal} {\bibinfo  {journal} {Phys. Rev. D}\ }\textbf {\bibinfo {volume}
  {78}},\ \bibinfo {pages} {053003} (\bibinfo {year} {2008})},\ \Eprint
  {http://arxiv.org/abs/0804.3627} {arXiv:0804.3627 [hep-ph]} \BibitemShut
  {NoStop}%
\bibitem [{\citenamefont {Gando}\ \emph {et~al.}(2016)\citenamefont {Gando}
  \emph {et~al.}}]{KamLAND-Zen:2016pfg}%
  \BibitemOpen
  \bibfield  {author} {\bibinfo {author} {\bibfnamefont {A.}~\bibnamefont
  {Gando}} \emph {et~al.} (\bibinfo {collaboration} {KamLAND-Zen}),\ }\href
  {\doibase 10.1103/PhysRevLett.117.082503} {\bibfield  {journal} {\bibinfo
  {journal} {Phys. Rev. Lett.}\ }\textbf {\bibinfo {volume} {117}},\ \bibinfo
  {pages} {082503} (\bibinfo {year} {2016})},\ \bibinfo {note} {[Addendum:
  Phys.Rev.Lett. 117, 109903 (2016)]},\ \Eprint
  {http://arxiv.org/abs/1605.02889} {arXiv:1605.02889 [hep-ex]} \BibitemShut
  {NoStop}%
\bibitem [{\citenamefont {Wong}(2011)}]{wong2011neutrino}%
  \BibitemOpen
  \bibfield  {author} {\bibinfo {author} {\bibfnamefont {Y.~Y.}\ \bibnamefont
  {Wong}},\ }\href@noop {} {\bibfield  {journal} {\bibinfo  {journal} {Annual
  Review of Nuclear and Particle Science}\ }\textbf {\bibinfo {volume} {61}},\
  \bibinfo {pages} {69} (\bibinfo {year} {2011})}\BibitemShut {NoStop}%
\bibitem [{\citenamefont {Aghanim}\ \emph {et~al.}(2020)\citenamefont {Aghanim}
  \emph {et~al.}}]{Aghanim:2018eyx}%
  \BibitemOpen
  \bibfield  {author} {\bibinfo {author} {\bibfnamefont {N.}~\bibnamefont
  {Aghanim}} \emph {et~al.} (\bibinfo {collaboration} {Planck}),\ }\href
  {\doibase 10.1051/0004-6361/201833910} {\bibfield  {journal} {\bibinfo
  {journal} {Astron. Astrophys.}\ }\textbf {\bibinfo {volume} {641}},\ \bibinfo
  {pages} {A6} (\bibinfo {year} {2020})},\ \Eprint
  {http://arxiv.org/abs/1807.06209} {arXiv:1807.06209 [astro-ph.CO]}
  \BibitemShut {NoStop}%
\bibitem [{\citenamefont {Mohapatra}\ and\ \citenamefont
  {Senjanovic}(1981)}]{Mohapatra:1980yp}%
  \BibitemOpen
  \bibfield  {author} {\bibinfo {author} {\bibfnamefont {R.~N.}\ \bibnamefont
  {Mohapatra}}\ and\ \bibinfo {author} {\bibfnamefont {G.}~\bibnamefont
  {Senjanovic}},\ }\href {\doibase 10.1103/PhysRevD.23.165} {\bibfield
  {journal} {\bibinfo  {journal} {Phys. Rev. D}\ }\textbf {\bibinfo {volume}
  {23}},\ \bibinfo {pages} {165} (\bibinfo {year} {1981})}\BibitemShut
  {NoStop}%
\bibitem [{\citenamefont {Gell-Mann}\ \emph {et~al.}(1979)\citenamefont
  {Gell-Mann}, \citenamefont {Ramond},\ and\ \citenamefont
  {Slansky}}]{GellMann:1980vs}%
  \BibitemOpen
  \bibfield  {author} {\bibinfo {author} {\bibfnamefont {M.}~\bibnamefont
  {Gell-Mann}}, \bibinfo {author} {\bibfnamefont {P.}~\bibnamefont {Ramond}}, \
  and\ \bibinfo {author} {\bibfnamefont {R.}~\bibnamefont {Slansky}},\
  }\href@noop {} {\bibfield  {journal} {\bibinfo  {journal} {Conf. Proc. C}\
  }\textbf {\bibinfo {volume} {790927}},\ \bibinfo {pages} {315} (\bibinfo
  {year} {1979})},\ \Eprint {http://arxiv.org/abs/1306.4669} {arXiv:1306.4669
  [hep-th]} \BibitemShut {NoStop}%
\bibitem [{\citenamefont {Yanagida}(1979)}]{Yanagida:1979as}%
  \BibitemOpen
  \bibfield  {author} {\bibinfo {author} {\bibfnamefont {T.}~\bibnamefont
  {Yanagida}},\ }\href@noop {} {\bibfield  {journal} {\bibinfo  {journal}
  {Conf. Proc. C}\ }\textbf {\bibinfo {volume} {7902131}},\ \bibinfo {pages}
  {95} (\bibinfo {year} {1979})}\BibitemShut {NoStop}%
\bibitem [{\citenamefont {Mohapatra}(1999)}]{Mohapatra:1999em}%
  \BibitemOpen
  \bibfield  {author} {\bibinfo {author} {\bibfnamefont {R.~N.}\ \bibnamefont
  {Mohapatra}},\ }\href@noop {} {\  (\bibinfo {year} {1999})},\ \Eprint
  {http://arxiv.org/abs/hep-ph/9910365} {arXiv:hep-ph/9910365} \BibitemShut
  {NoStop}%
\bibitem [{\citenamefont {Ashtari~Esfahani}\ \emph {et~al.}(2017)\citenamefont
  {Ashtari~Esfahani} \emph {et~al.}}]{Esfahani:2017dmu}%
  \BibitemOpen
  \bibfield  {author} {\bibinfo {author} {\bibfnamefont {A.}~\bibnamefont
  {Ashtari~Esfahani}} \emph {et~al.} (\bibinfo {collaboration} {Project 8}),\
  }\href {\doibase 10.1088/1361-6471/aa5b4f} {\bibfield  {journal} {\bibinfo
  {journal} {J. Phys. G}\ }\textbf {\bibinfo {volume} {44}},\ \bibinfo {pages}
  {054004} (\bibinfo {year} {2017})},\ \Eprint
  {http://arxiv.org/abs/1703.02037} {arXiv:1703.02037 [physics.ins-det]}
  \BibitemShut {NoStop}%
\bibitem [{\citenamefont {Shirai}(2017)}]{Shirai:2017jyz}%
  \BibitemOpen
  \bibfield  {author} {\bibinfo {author} {\bibfnamefont {J.}~\bibnamefont
  {Shirai}} (\bibinfo {collaboration} {KamLAND-Zen}),\ }\href {\doibase
  10.1088/1742-6596/888/1/012031} {\bibfield  {journal} {\bibinfo  {journal}
  {J. Phys. Conf. Ser.}\ }\textbf {\bibinfo {volume} {888}},\ \bibinfo {pages}
  {012031} (\bibinfo {year} {2017})}\BibitemShut {NoStop}%
\bibitem [{\citenamefont {Giuliani}\ \emph {et~al.}(2019)\citenamefont
  {Giuliani} \emph {et~al.}}]{Giuliani:2019uno}%
  \BibitemOpen
  \bibfield  {author} {\bibinfo {author} {\bibfnamefont {A.}~\bibnamefont
  {Giuliani}} \emph {et~al.} (\bibinfo {collaboration} {APPEC Committee}),\
  }\href@noop {} {\  (\bibinfo {year} {2019})},\ \Eprint
  {http://arxiv.org/abs/1910.04688} {arXiv:1910.04688 [hep-ex]} \BibitemShut
  {NoStop}%
\bibitem [{\citenamefont {Brinckmann}\ \emph {et~al.}(2019)\citenamefont
  {Brinckmann}, \citenamefont {Hooper}, \citenamefont {Archidiacono},
  \citenamefont {Lesgourgues},\ and\ \citenamefont
  {Sprenger}}]{Brinckmann:2018owf}%
  \BibitemOpen
  \bibfield  {author} {\bibinfo {author} {\bibfnamefont {T.}~\bibnamefont
  {Brinckmann}}, \bibinfo {author} {\bibfnamefont {D.~C.}\ \bibnamefont
  {Hooper}}, \bibinfo {author} {\bibfnamefont {M.}~\bibnamefont
  {Archidiacono}}, \bibinfo {author} {\bibfnamefont {J.}~\bibnamefont
  {Lesgourgues}}, \ and\ \bibinfo {author} {\bibfnamefont {T.}~\bibnamefont
  {Sprenger}},\ }\href {\doibase 10.1088/1475-7516/2019/01/059} {\bibfield
  {journal} {\bibinfo  {journal} {JCAP}\ }\textbf {\bibinfo {volume} {01}},\
  \bibinfo {pages} {059} (\bibinfo {year} {2019})},\ \Eprint
  {http://arxiv.org/abs/1808.05955} {arXiv:1808.05955 [astro-ph.CO]}
  \BibitemShut {NoStop}%
\bibitem [{\citenamefont {Alion}\ \emph {et~al.}(2016)\citenamefont {Alion}
  \emph {et~al.}}]{DUNE:2016ymp}%
  \BibitemOpen
  \bibfield  {author} {\bibinfo {author} {\bibfnamefont {T.}~\bibnamefont
  {Alion}} \emph {et~al.} (\bibinfo {collaboration} {DUNE}),\ }\href@noop {} {\
   (\bibinfo {year} {2016})},\ \Eprint {http://arxiv.org/abs/1606.09550}
  {arXiv:1606.09550 [physics.ins-det]} \BibitemShut {NoStop}%
\bibitem [{\citenamefont {Abe}\ \emph {et~al.}(2018)\citenamefont {Abe} \emph
  {et~al.}}]{Hyper-Kamiokande:2018ofw}%
  \BibitemOpen
  \bibfield  {author} {\bibinfo {author} {\bibfnamefont {K.}~\bibnamefont
  {Abe}} \emph {et~al.} (\bibinfo {collaboration} {Hyper-Kamiokande}),\
  }\href@noop {} {\  (\bibinfo {year} {2018})},\ \Eprint
  {http://arxiv.org/abs/1805.04163} {arXiv:1805.04163 [physics.ins-det]}
  \BibitemShut {NoStop}%
\bibitem [{\citenamefont {Froggatt}\ and\ \citenamefont
  {Nielsen}(1979)}]{froggatt1979hierarchy}%
  \BibitemOpen
  \bibfield  {author} {\bibinfo {author} {\bibfnamefont {C.~D.}\ \bibnamefont
  {Froggatt}}\ and\ \bibinfo {author} {\bibfnamefont {H.~B.}\ \bibnamefont
  {Nielsen}},\ }\href@noop {} {\bibfield  {journal} {\bibinfo  {journal}
  {Nuclear Physics B}\ }\textbf {\bibinfo {volume} {147}},\ \bibinfo {pages}
  {277} (\bibinfo {year} {1979})}\BibitemShut {NoStop}%
\bibitem [{\citenamefont {Nir}(2016)}]{nir2016flavour}%
  \BibitemOpen
  \bibfield  {author} {\bibinfo {author} {\bibfnamefont {Y.}~\bibnamefont
  {Nir}},\ }\href@noop {} {\bibfield  {journal} {\bibinfo  {journal} {arXiv
  preprint arXiv:1605.00433}\ } (\bibinfo {year} {2016})}\BibitemShut {NoStop}%
\bibitem [{\citenamefont {Hall}\ \emph {et~al.}(2000)\citenamefont {Hall},
  \citenamefont {Murayama},\ and\ \citenamefont {Weiner}}]{Hall:1999sn}%
  \BibitemOpen
  \bibfield  {author} {\bibinfo {author} {\bibfnamefont {L.~J.}\ \bibnamefont
  {Hall}}, \bibinfo {author} {\bibfnamefont {H.}~\bibnamefont {Murayama}}, \
  and\ \bibinfo {author} {\bibfnamefont {N.}~\bibnamefont {Weiner}},\ }\href
  {\doibase 10.1103/PhysRevLett.84.2572} {\bibfield  {journal} {\bibinfo
  {journal} {Phys. Rev. Lett.}\ }\textbf {\bibinfo {volume} {84}},\ \bibinfo
  {pages} {2572} (\bibinfo {year} {2000})},\ \Eprint
  {http://arxiv.org/abs/hep-ph/9911341} {arXiv:hep-ph/9911341} \BibitemShut
  {NoStop}%
\bibitem [{\citenamefont {Haba}\ and\ \citenamefont
  {Murayama}(2001)}]{Haba:2000be}%
  \BibitemOpen
  \bibfield  {author} {\bibinfo {author} {\bibfnamefont {N.}~\bibnamefont
  {Haba}}\ and\ \bibinfo {author} {\bibfnamefont {H.}~\bibnamefont
  {Murayama}},\ }\href {\doibase 10.1103/PhysRevD.63.053010} {\bibfield
  {journal} {\bibinfo  {journal} {Phys. Rev. D}\ }\textbf {\bibinfo {volume}
  {63}},\ \bibinfo {pages} {053010} (\bibinfo {year} {2001})},\ \Eprint
  {http://arxiv.org/abs/hep-ph/0009174} {arXiv:hep-ph/0009174} \BibitemShut
  {NoStop}%
\bibitem [{\citenamefont {de~Gouvea}\ and\ \citenamefont
  {Murayama}(2003)}]{deGouvea:2003xe}%
  \BibitemOpen
  \bibfield  {author} {\bibinfo {author} {\bibfnamefont {A.}~\bibnamefont
  {de~Gouvea}}\ and\ \bibinfo {author} {\bibfnamefont {H.}~\bibnamefont
  {Murayama}},\ }\href@noop {} {\bibfield  {journal} {\bibinfo  {journal}
  {Phys. Lett. B}\ }\textbf {\bibinfo {volume} {573}},\ \bibinfo {pages} {94}
  (\bibinfo {year} {2003})},\ \Eprint {http://arxiv.org/abs/hep-ph/0301050}
  {arXiv:hep-ph/0301050} \BibitemShut {NoStop}%
\end{thebibliography}%

\end{document}